\documentclass[10pt,english]{article}
\usepackage{geometry}
\usepackage{setspace}
\usepackage[T1]{fontenc}
\usepackage[utf8]{inputenc}
\usepackage{babel}
\usepackage{varioref}
\usepackage{amsmath}
\numberwithin{equation}{section}
\usepackage{amssymb}
\usepackage{upgreek}
\usepackage{graphicx}
\usepackage{floatrow}
\usepackage[labelfont=bf]{caption}
\usepackage{subcaption}
\usepackage[shortlabels]{enumitem}
\usepackage[toc,page]{appendix}
\usepackage[normalem]{ulem}
\newcommand{\nocontentsline}[3]{}
\newcommand{\tocless}[2]{\bgroup\let\addcontentsline=\nocontentsline#1{#2}\egroup}

\usepackage{slashed}
\usepackage{svg}
\usepackage [autostyle, english = american]{csquotes}
\MakeOuterQuote{"}
\makeatletter

\usepackage{xcolor}
\usepackage{authblk}

\usepackage{amsthm}
\usepackage{graphicx}

\usepackage{mathtools}
\usepackage{comment}

\usepackage{xargs}                      

\usepackage[colorinlistoftodos,prependcaption,textsize=tiny]{todonotes}
\newcommandx{\unsure}[2][1=]{\todo[linecolor=red,backgroundcolor=red!25,bordercolor=red,#1]{#2}}
\newcommandx{\change}[2][1=]{\todo[linecolor=blue,backgroundcolor=blue!25,bordercolor=blue,#1]{#2}}
\newcommandx{\info}[2][1=]{\todo[linecolor=green,backgroundcolor=green!25,bordercolor=green,#1]{#2}}
\newcommandx{\improve}[2][1=]{\todo[linecolor=violet,backgroundcolor=violet!25,bordercolor=violet,#1]{#2}}
\newcommandx{\thiswillnotshow}[2][1=]{\todo[disable,#1]{#2}}
\usepackage{hyperref}
\hypersetup{
    colorlinks,
    linkcolor={blue!60!black},
    citecolor={green!60!black},
    urlcolor={blue!80!black}
}
\newtheorem{thm}{Theorem}[section]

\theoremstyle{remark}

\theoremstyle{plain}

\theoremstyle{remark}
\usepackage{todonotes}

\newcommand{\dd}{\mathop{}\!\mathrm{d}}

\newcommand{\pu}{\partial_u}
\newcommand{\pv}{\partial_v}

\title{On the Relation Between Asymptotic Charges,\\ the Failure of Peeling and Late-time Tails} 

\author[1]{Dejan Gajic\thanks{dejan.gajic@uni-leipzig.de}}
\author[2]{Lionor M. A. Kehrberger\thanks{kehrberger@mis.mpg.de}} 

\affil[1]{Radboud University, Department of Mathematics, Huygens Building,
Heyendaalseweg 135,
6525 AJ Nijmegen,
The Netherlands}
\affil[2]{University of Cambridge, Department of Applied Mathematics and Theoretical Physics, Wilberforce Road, Cambridge CB3 0WA, United Kingdom}
\date{October 2, 2023} 

\makeatother

\begin{document}
\pagenumbering{gobble}

\maketitle 
\begin{abstract}
The last few years have seen considerable mathematical progress concerning the asymptotic structure of gravitational radiation in dynamical, astrophysical spacetimes. 

In this paper, we distil some of the key ideas from recent works and assemble them in a new way in order to make them more accessible to the wider general relativity community. In the process, we also discuss new physical findings.

First, we introduce the conserved~$f(r)$-modified Newman--Penrose charges on asymptotically flat spacetimes, and we show that these charges provide a dictionary that relates asymptotics of massless, general spin fields in different regions: Asymptotic behaviour near~$i^+$ ("late-time tails") can be read off from asymptotic behaviour towards~$\mathcal I^+$, and, similarly, asymptotic behaviour towards~$\mathcal I^+$ can be read off from asymptotic behaviour near~$i^-$ or~$\mathcal I^-$.

Using this dictionary, we then explain how: \textbf{(I)} the quadrupole approximation for a system of~$N$ infalling masses from~$i^-$ causes the "peeling property towards~$\mathcal I^+$" to be  violated, and \textbf{(II)} this failure of peeling results in deviations from the usual predictions for tails in the late-time behaviour of gravitational radiation:
Instead of the Price's law rate~$r\Psi^{[4]}|_{\mathcal I^+}\sim  u^{-6}$ as~$u\to\infty$, we predict that~$r\Psi^{[4]}|_{\mathcal I^+}\sim  u^{-3}$, with the coefficient of this latter decay rate being a multiple of the monopole and quadrupole moments of the matter distribution in the infinite past.
\end{abstract}

\begingroup
\hypersetup{linkcolor=black}
\setcounter{tocdepth}{1}
    \tableofcontents
    \endgroup

\section{Introduction}
\pagenumbering{arabic}
Over the last few years, there has been significant progress in the mathematical study of general relativity concerning our understanding of how certain asymptotic conservation laws---related to the \textit{Newman--Penrose charges} \cite{NPconstants65,NPconstants68}---can serve as a mechanism for deriving statements about asymptotics of gravitational waves on black hole spacetimes. 
This mechanism was first explored in~\cite{AAG18b} and has since lead to new, potentially physically measurable predictions~\cite{AAGLetters, burko19, I,II}.
The present paper aims to distil and generalise some of the main ideas behind this mechanism, and to provide a physical and more accessible interpretation thereof.
We also preview several upcoming results.

The ideas presented in this note generally pertain to the system of linearised equations of gravity around a Kerr black hole background~$(\mathcal M_{M,a},g_{M,a})$. This system of equations both contains, and is, at a fundamental level, governed\footnote{In particular, any admissible perturbation that has $\Psi^{[0]}=0=\Psi^{[4]}$ must be the sum of a pure gauge solution and a linearised Kerr solution \cite{WaldTeuk}.} by the spin-$\pm2$ Teukolsky equations\footnote{In fact, the ideas discussed in the present paper apply to the Teukolsky equations \textit{for any spin}~$s\in \frac{1}{2}\mathbb{N}_0$.}~\cite{Teukolsky73}: 
\begin{equation}\label{eq:Teukolsky}
\mathcal T_{g_{M,a}}^{[s]} \Psi^{[|s|- s]}=0, \quad s=\pm 2,
\end{equation} 
with~$\mathcal T_{g_{M,a}}^{[s]}$ a differential operator similar to the wave operator~$\Box_{g_{M,a}}$ on Kerr, and with $\Psi^{[0]}$ and $\Psi^{[4]}$ the gauge-invariant, extremal components of the perturbed Weyl tensor in the Newman--Penrose formalism~\cite{NP62Approach}. 

To keep the presentation as clear as possible, however, we will restrict most of the discussion to the simpler case of~$s=0$, in which \eqref{eq:Teukolsky} in fact equals the scalar wave equation
\begin{equation}\label{eq:wave}
\Box_{g_{M,a}} \psi=0.
\end{equation}
We will moreover restrict to the subcase with specific angular momentum~$a=0$, where~$g_{M,a}$ reduces to the Schwarzschild metric~$g_{M}$. 
Extensions to non-zero~$s$ and~$a$ will be discussed at the end of the paper.

\begin{figure}[htpb]
\includegraphics[width=120pt]{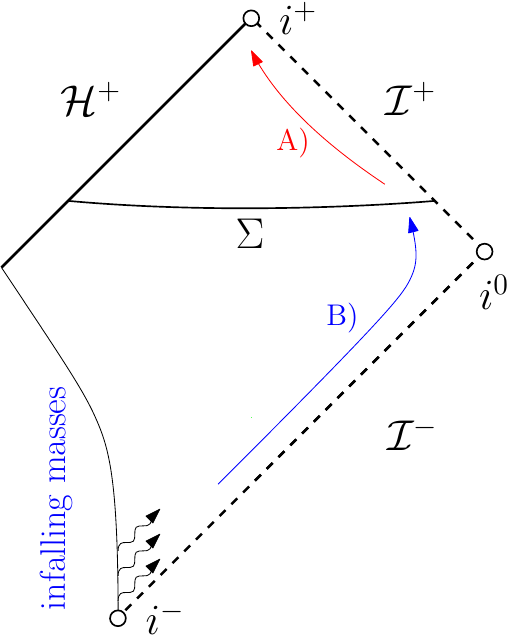}
\caption{Depiction of the problems \textbf{A)} and \textbf{B)} described in the paragraph above.}
\label{fig:intro}
\end{figure}

This paper will answer the following two questions (see Fig.~\ref{fig:intro}): 
\begin{itemize}
\item[\textbf{A)}] How can we read off late-time asymptotics of~$\psi$ near future timelike infinity~$i^+$---in particular, \textit{along} future null infinity~$\mathcal I^+$ and the event horizon~$\mathcal H^+$---from asymptotics \textit{towards}~$\mathcal I^+$?
\item[\textbf{B)}] How can we derive asymptotics towards~$\mathcal I^+$ from physically motivated scattering data assumptions modelling a system of~$N$ infalling masses from the infinite past~$i^-$ and excluding incoming radiation from past null infinity~$\mathcal I^-$?
\end{itemize}

\subsection{Motivation and background}
Let us first motivate problems \textbf{A)} and \textbf{B)} individually:

\textbf{A)} 
The study of the dynamics of \eqref{eq:Teukolsky} at late times is motivated by the ambitious final state conjecture, which in particular asserts that two inspiralling black holes will settle down to a Kerr solution outside the horizon~\cite{Penrose69collapse},\footnote{In the restricted case of spacetimes arising from small perturbations of subextremal Kerr initial data, this conjecture is known as the Kerr stability conjecture, see for example~\cite{DHRT21, KlSz21} and references therein for recent mathematical progress towards a proof thereof.} and by the fact that, mathematically, a large part of the radiation emitted as dynamical spacetimes settle down to a Kerr spacetime is encoded precisely in the solutions~$\Psi^{[4]}$ and~$\Psi^{[0]}$ to eq.\ \eqref{eq:Teukolsky}. 

Hence, in the idealisation of an isolated gravitational system, where gravitational wave observatories operate at~$\mathcal I^+$, the mathematical late-time behaviour (of appropriate rescalings) of the Teukolsky variables~$\Psi^{[4]}$ and~$\Psi^{[0]}$ along~$\mathcal I^+$ offers predictions for the late-time part of signals measured by actual gravitational wave detectors; see for example the analysis of the late-time parts of gravitational signals coming from recent black hole mergers in~\cite{ligotest, LIGOlatetime}. 
This gives rise to several interesting points:
\begin{enumerate}[1)]
\item 
Suppose one manages to measure the decay rates and coefficients in the late-time asymptotics of the Teukolsky variables, and interprets these as asymptotics towards~$i^+$ along~$\mathcal I^+$. Given a good mathematical understanding of these late-time tails, one may then extract from the measurements information about both the nature of the final Kerr black hole state as well as the asymptotic properties of the initial state.
Late-time tails therefore serve as important signatures of black holes in dynamical astrophysical processes. 
\item In addition, the late-time behaviour along~$\mathcal{I}^+$ is \emph{mathematically strongly correlated} with the late-time behaviour of appropriate renormalisations of the Teukolsky variables along~$\mathcal{H}^+$. 
\item In turn, the late-time behaviour along~$\mathcal{H}^+$ is related to the strength of the null singularity that is expected to exist generically inside dynamical black holes; see for example~\cite{Dafermos05interior, LukOh16}. 
A sufficiently precise understanding of the behaviour of the Teukolsky variables along~$\mathcal H^+$ is therefore necessary for resolving the \emph{Strong Cosmic Censorship conjecture} (see~\cite{DafermosLuk17} and references therein) in the setting of dynamical black hole spacetimes.
Hence, via 2), the information measured along~$\mathcal I^+$ would also relate to behaviour in the black hole interior!
\end{enumerate}

Of course, the late-time asymptotics near~$i^+$ will depend on the assumptions made on the initial state, i.e.\ the choice of initial data hypersurface and the prescribed initial data. 
For instance, if the initial data are posed on an asymptotically hyperboloidal hypersurface~$\Sigma$ (see Fig.~\ref{fig:intro}), 
we will see that the leading-order asymptotics near~$i^+$ are, in many cases, completely determined by how the data for~$\psi$ behave near~$\mathcal I^+$, i.e.\ by the asymptotics of~$\psi$ along~$\Sigma$ towards~$\mathcal I^+$.
But what should these {early-time} asymptotics towards~$\mathcal I^+$  be?

\textbf{B)} In large (but not all) parts of the literature, there have been two predominant data assumptions on the asymptotics towards~$\mathcal I^+$ along $\Sigma$:
It has been assumed (e.g.\ in the original heuristic work on late-time asymptotics~\cite{Price72}) that these asymptotics are either trivial, i.e.\ that the data are of compact support along~$\Sigma$ and therefore vanish identically near~$\mathcal I^+$, or---typically justified by Penrose's concept of smooth conformal compactification of spacetime (a.k.a.\ smooth null infinity)~\cite{Penrose65}---it has been assumed that the initial data satisfy "peeling" \cite{SeriesVI,SeriesVIII}, i.e.\ that they have an asymptotic expansion in powers of~$1/r$ and exhibit certain leading-order decay towards~$\mathcal I^+$ (e.g.~$\Psi^{[0]}=O(r^{-5}), \Psi^{[4]}=\mathcal O(r^{-1})$).

Now, we would argue that the former assumption is incompatible with the model of an isolated system, as any such system will have radiated \textit{for all times} and, therefore, will not have hypersurfaces~$\Sigma$ of compact radiation content, see Fig.~\ref{fig:intro}.

The assumption of peeling becomes similarly questionable for physically relevant systems, as will be explained in this paper.
Indeed, the approach we take here is to \textit{not make assumptions} on the asymptotics along $\Sigma$ towards $\mathcal I^+$, but to instead \textit{derive them from physical principles}: 
We shall consider a \textit{scattering data} setup as in~\cite{I} that \textbf{a)} has no incoming radiation from~$\mathcal I^-$ and that \textbf{b)} attempts to capture the gravitational radiation of~$N$  masses---approaching each other from infinitely far away in the infinite past---by imposing data on some null cone~$\mathcal C$ (to be thought of as enclosing these masses) that are predicted by post-Newtonian arguments (such as the quadrupole approximation) \cite{WalkerWill79,Damour86,Chr02}.
From this scattering setup, we will then \textit{dynamically derive} the asymptotics towards~$\mathcal I^+$, see Fig.~\ref{fig:intro:thm}.
\subsection{The main result}
\begin{figure}[htpb]
\includegraphics[width=130pt]{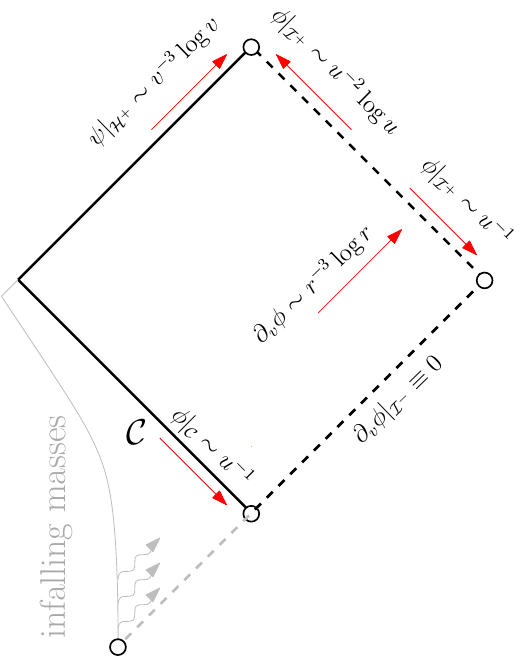}
\caption{Summary of the main results of the paper, focussing on the spherically symmetric ($\ell=0$) part of the scalar field ($s=0$).}\label{fig:intro:thm}
\end{figure}

Let us here already give an outline of the results we obtain from the scattering data setup described above, focussing first on the simpler case $s=0$. See also Fig.~\ref{fig:intro:thm}.
\begin{itemize}
\item We start by assuming that, if~$\psi_0$ denotes the spherically symmetric part of~$\psi$, and if~$\phi_0:=r\psi_0$, then~$\phi_0|_{\mathcal C}\sim u^{-1}$ as~$u\to-\infty$ along some ingoing null hypersurface~$\mathcal C$. This assumption is motivated by post-Newtonian arguments for systems of $N$ infalling masses from $i^-$, see \S\ref{sec:goodmorningtales}.
\item Combining this assumption with the condition of no incoming radiation from~$\mathcal I^-$, we then prove that~$\pv\phi_0\sim r^{-3}\log r$ towards~$\mathcal I^+$, so the peeling property fails. In fact, we prove that the limit $\lim_{\mathcal I^+}\frac{r^3}{\log r} \pv\phi_0$ is conserved along $\mathcal I^+$.
\item Finally, if one smoothly, but arbitrarily, extends the data along~$\mathcal C$ towards~$\mathcal H^+$, then this failure of peeling will lead to the following late-time asymptotics \textit{along}~$\mathcal I^+$:~$\phi_0|_{\mathcal I^+}\sim u^{-2}\log u$ as~$u\to\infty$. 
This should be contrasted with the Price's law\footnote{Price's law is the statement that the following asymptotic behaviour should hold given sufficiently rapidly decaying data:~$\psi_{\ell}\sim t^{-3-2\ell}$ along curves of constant~$r$ as~$t\to \infty$, and~$\phi_{\ell}|_{\mathcal I^+}\sim u^{-2-\ell}$ as~$u\to \infty$. These asymptotics were first predicted from heuristic arguments in~\cite{Price72} and~\cite{Leaver86}, respectively, and proved using mathematically rigorous arguments in~\cite{AAG18b, Hintz22, AAG21}.} rate one obtains for Cauchy data of compact support:~$\phi_0|_{\mathcal I^+}\sim u^{-2}$.
\end{itemize}

For the scalar field ($s=0$), the \textit{measurable} rate along~$\mathcal I^+$ thus only differs by a logarithm from the Price's law rate.
This difference is much less subtle in the case of gravitational perturbations ($|s|=2$):
\textbf{There, under the same setup, we expect the following:~$\Psi^{[0]}$ violates the peeling property towards~$\mathcal I^+$,~$\Psi^{[0]}\sim \mathcal O(r^{-4})$, and this leads to~$r\Psi^{[4]}|_{\mathcal I^+}$ decaying like~$u^{-3}$ along~$\mathcal I^+$ as~$u\to\infty$, as opposed to the~$u^{-6}$-rate that one obtains in the case of compactly supported Cauchy data  and the~$u^{-5}$-rate that one obtains for data consistent with the peeling property!}

The proof of this expectation is work in progress~\cite{IV,V}, but \S\ref{sec:dictionary} of the present paper already contains a preview of the argument.
\subsection{Structure}
The geometry, coordinates and foliations of the Schwarzschild spacetime are set up in \S\ref{sec:setupGK}.
We then give the definitions of the~$f(r)$-modified Newman--Penrose charges and their associated conservation laws in \S\ref{sec:NP}.
In~\S\ref{sec:goodnighttales}, we use these conservation laws to explain how to translate various asymptotics  \textit{towards~$\mathcal I^+$} into late-time asymptotics \textit{along~$\mathcal I^+$} and near~$i^+$.
In~\S\ref{sec:goodmorningtales}, we then construct a simple model capturing a system of~$N$ infalling masses from~$i^-$ and discuss the asymptotics towards~$\mathcal I^+$ exhibited by this model.
In~\S\ref{sec:dictionary}, we combine the results of \S\ref{sec:goodmorningtales} and \S\ref{sec:goodnighttales} to obtain a complete dictionary translating asymptotics near~$\mathcal I^-$ to asymptotics near~$i^+$. In particular, this gives predictions on the (in principle) measurable late-time asymptotics along~$\mathcal I^+$.

We briefly touch upon some extensions of the methods to different settings in \S\ref{sec:Extending}, and we conclude in \S\ref{sec:conclusions}.

\section{Coordinates, foliations and conventions}\label{sec:setupGK}
We consider the Schwarzschild black hole exterior spacetimes, denoted~$(\mathring{\mathcal{M}}_M,g_M)$, where~$\mathring{\mathcal{M}}_M=\mathbb{R}_t\times (2M,\infty)_r\times \mathbb{S}^2_{(\theta,\varphi)}$, and where
\begin{align*}
g_M=-D\dd t^2+D^{-1}\dd r^2+r^2(\dd \theta^2+\sin^2\theta \dd \varphi^2), \quad D=1-\tfrac{2M}{r}.
\end{align*}

\begin{figure}[htpb]
\includegraphics[width=150pt]{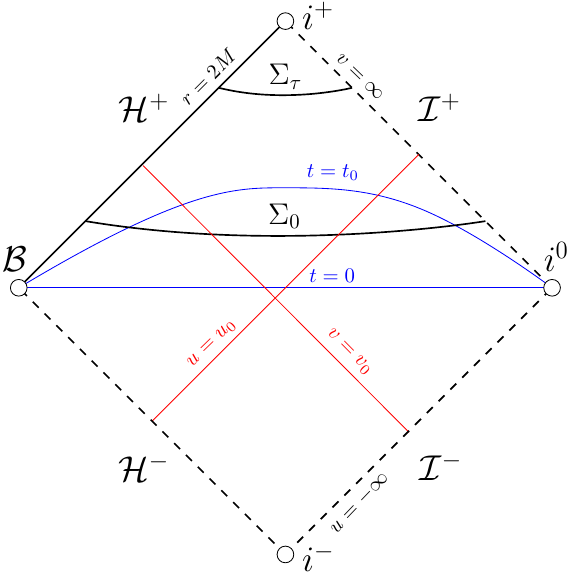}
\caption{A Penrose diagrammatic depiction of~$\mathcal{M}_M$ with the asymptotically hyperboloidal foliation by~$\Sigma_{\tau}$, constant-$t$ level sets and~$u$- and~$v$-null hypersurfaces.}\label{fig:foliations}
\end{figure}

Note that the constant-$t$ slices foliating~$\mathring{\mathcal{M}}_M$ are \emph{asymptotically flat} and, in the Kruskal extension of the spacetime, approach the bifurcation sphere~$\mathcal B$ (see Fig.~\ref{fig:foliations}).
To capture radiative properties in the spacetime, it will be more convenient to introduce the following~$\tau$-slicing by \emph{asymptotically hyperboloidal} hypersurfaces that penetrate the event horizon strictly to the future of $\mathcal B$: 
Consider the new time function
\begin{align*}
\tau=t+r_*-2(r-2M)-4M \log \left(\tfrac{r}{2M}\right)\quad \textnormal{with}\quad
r_*=r+2M\log \left(\tfrac{r-2M}{M}\right).
\end{align*}
It may easily be verified that we have the following expression for~$g_M$ in $(\tau,r,\theta,\varphi)$ coordinates:
\begin{equation*}
g_M=-(1-\tfrac{2M}{r})\dd\tau^2-2(1-\tfrac{8M^2}{ r^{2}})\dd\tau \dd r+\tfrac{16M^2}{r^2}(1+\tfrac{2M}{r})\dd r^2+r^2(\dd\theta^2+\sin^2\theta \dd\varphi^2).
\end{equation*}
In fact, this metric is  well-defined on the manifold-with-boundary~$\mathcal{M}_M=\mathbb{R}_{\tau}\times [2M,\infty)_r\times \mathbb{S}^2_{(\theta,\varphi)}$, which may be thought of as an extension of~$\mathring{\mathcal{M}}_M$ that includes the future event horizon~$\mathcal{H}^+$ as the level set~$\mathcal{H}^+=\{r=2M\}$.
We denote with~$\Sigma_{\tau}$ the constant-$\tau$ level sets:
\begin{equation*}
\Sigma_{\tau_0}=\{\tau=\tau_0\}.
\end{equation*}

We will moreover make use of the double null functions
\begin{align}\label{eq:doublenull}
v=\:t+r_*,&&
u=\:t-r_*,
\end{align}
and we will consider double null coordinates~$(u,v,\theta,\varphi)$ on~$\mathring{\mathcal{M}}_M$. 
The level sets of constant~$u$ or~$v$ are null hypersurfaces, 
and we may formally\footnote{One can also view these sets as conformal boundaries of the spacetime.} represent future and past null infinity,~$\mathcal{I}^+$ and~$\mathcal I^-$, as the level sets
\begin{align*}
\mathcal{I}^+=\{v=\infty\},&&\mathcal I^-=\{u=-\infty\}.
\end{align*}

A final word on conventions: The letter~$C$ appearing in inequalities will always denote a positive constant whose precise value is not relevant and which is allowed to change from line to line. 
In other words,~$C$ obeys the "algebra of constants":~$C\cdot C=C+C=C\dots$. We will occasionally omit~$C$ and just write~$f\lesssim g$ or~$f\gtrsim g$ if the quantities~$f,g$ satisfy~$f\leq C g$ or~$f \geq C g$, respectively. If both~$f \lesssim g$ and~$f \gtrsim g$, we write~$f\sim g$.

If we talk about constants whose specific values do matter, we will typically use either the letter~$A$ or~$B$, the former in the case when~$A$ is just a constant and nonvanishing multiple of certain data quantities, the latter when~$B$ is a more complicated expression (that can in principle vanish).

\section{The~$f(r)$-modified Newman--Penrose charges}\label{sec:NP}
In double null coordinates \eqref{eq:doublenull}, one can derive from \eqref{eq:wave} the following wave equation for the rescaled quantity~$\phi:=r\psi$:
\begin{equation}\label{eq:NP:wave}
4\pu\pv\phi=\frac{D}{r^2}{\slashed{\Delta}_{\mathbb S^2}\phi}-\frac{2MD}{r^3}\phi,
\end{equation}
where~$\slashed{\Delta}_{\mathbb S^2}$ is the spherical Laplacian with eigenvalues~${-\ell(\ell+1)}$,~$\ell\in\mathbb N_0$. 

To develop an intuition for the~$f(r)$-modified Newman--Penrose (N--P) charges, let us first consider\textbf{ the case~$M=0$:}
Projecting onto spherical harmonics~$Y_{\ell m}$,~$\phi_{\ell m}:=\langle\phi,Y_{\ell m}\rangle_{L^2(\mathbb S^2)}$ (and suppressing the~$m$-index), it is straightforward to derive from \eqref{eq:NP:wave} the following infinite set of \textit{exact conservation laws} for $N=\ell$:
\begin{equation}\label{eq:NP:NPMink}
\pu(r^{-2N}\pv(r^2\pv)^N\phi_{\ell})=\frac{(N-\ell)(N+\ell+1)}{r^{2N+2}}(r^2\pv)^N\phi_{\ell}.
\end{equation}

If~$M\neq0$, we no longer have global conservation laws, but we can still derive \textit{asymptotic} conservation laws. 
\textbf{Focussing first on the~$\ell=0$-mode~$\phi_0$,} the analogue of \eqref{eq:NP:NPMink}  reads
\begin{equation}\label{eq:NP:NPl=0}
4\pu\pv \phi_0=-\frac{2MD}{r^3}\phi_0.
\end{equation}
Clearly,~$\pv\phi_0$ is no longer globally conserved; however, the RHS exhibits a good $r^{-3}$-weight. Therefore, for any function~$f$ with 
\begin{equation}
(f(r))^{-1}=o(r^{3}),
\end{equation}
 the \textit{$f(r)$-modified Newman--Penrose charge}
\begin{equation}
I_0^f[\psi](u):=\lim_{v\to\infty}(f(r))^{-1}\pv\phi_0(u,v)
\end{equation}
will, under suitable assumptions and so long as it exists for some value of~$u$,  be conserved in~$u$.
This essentially follows from commuting \eqref{eq:NP:NPl=0} with~$f(r)^{-1}$, see eq.~\eqref{eq:step1} in \S\ref{sec:nonzeroNP} for a proof.

While the charges $I_0^f[\psi]$ have first been defined for~$f(r)=r^{-2}$~\cite{NPconstants68, AAG21}, we will see in the present paper that other choices of~$f$ are equally important; see also~\cite{Kroon00,Kroon01,II,III}.

Notice however the restriction~$f(r)^{-1}=o(r^3)$ (also a consequence of the~$r^{-3}$-weight in \eqref{eq:NP:NPl=0}). 
In particular, if~$\pv\phi_0=O(r^{-3})$ initially, then \textit{all} ($\ell=0$) $f(r)$-modified N--P charges vanish, and one cannot \textit{directly} associate a non-zero asymptotic conserved charge to \eqref{eq:NP:NPl=0}. 

\newcommand{\VV}{\boldsymbol{\widehat{L} }}
\paragraph{Generalising \eqref{eq:NP:NPl=0} to higher~$\ell$:} If we naively compute the RHS of \eqref{eq:NP:NPMink} for $M\neq 0$ and $N=\ell$, then the highest-order term in derivatives will be adorned with a good $r^{-2\ell-3}$-weight, whereas all lower-order derivatives will come with a bad $r^{-2\ell-2}$-weight. This problem can be addressed by considering not $(r^2\pv)^\ell\phi_\ell$, but a suitable combination of $(r^2\pv)^\ell\phi_\ell$ and lower-order derivatives. Moreover, it is more natural to work with the rescaled null vector field~$\VV :=D^{-1}r^2\pv$ rather than with $r^2\pv$.\footnote{Note that in coordinates~$(u,x=1/r, \theta, \varphi)$, we have~$2\VV =-\partial_x$.\label{prevft}} To be precise, if we replace~$(r^2\pv)^\ell \phi_\ell$ in \eqref{eq:NP:NPMink} with 
\begin{equation}\label{PHIL}
\Phi_\ell:=\VV ^\ell\phi_\ell+\sum_{i=1}^\ell x_i^{(\ell)}\cdot M^i \cdot \VV ^{\ell-i}\phi_\ell, \quad \text{where}\quad x_i^{(\ell)}=\frac{1}{i!}\frac{(2\ell-i)!}{(2\ell)!}\left(\frac{\ell!}{(\ell-i)!}\right)^3,
\end{equation}
then a lengthy computation shows that
\begin{equation}\label{eq:NP:NPl}
\pu(D^\ell r^{-2\ell}\pv \Phi_\ell)=\frac{D^{\ell+1}M}{r^{2\ell+3}}\sum_{i=0}^\ell \left(y_i^{(\ell)}+\frac{M}{r}z_i^{(\ell)}\right)\cdot M^i\cdot  \VV ^{\ell-i}\phi_\ell ,
\end{equation}
where~$\{y_i^{(\ell)}, z_i^{(\ell)}\}$ is a set of numerical constants.
As before, we can now associate, for any~$f$ with~$(f(r))^{-1}=o(r^3)$, the following~$f(r)$-modified N--P charges to \eqref{eq:NP:NPl}:
\begin{equation}
I^f_\ell[\psi](u):=\lim_{v\to\infty}f(r)^{-1}\pv \Phi_\ell(u,v).
\end{equation}
Under suitable assumptions, these~$I^f_\ell[\psi](u)$ will again be conserved if finite for some value of~$u$, see \S\ref{sec:nonzeroNP}.

Importantly, in addition to being conserved, the charges~$I^f_\ell[\psi]$ also provide a measure of conformal regularity, i.e.\ regularity in the variable~$x=1/r$, of the field~$\psi$, see Footnote~\ref{prevft}.
In order to illustrate this point, we consider the following example:
Given data~$\psi$ on the hyperboloidal hypersurface~$\Sigma_0$ that only have a finite asymptotic expansion in powers of~$1/r$, say,~$$r\psi_\ell=\sum_{i=0}^{\lceil p_\ell\rceil}\frac{C_i}{r^{i}}+C^\ast \frac{ \log^{q_\ell} r}{r^{p_\ell+1}}+\dots \quad \text{for some } q_\ell\in\mathbb N, p_\ell\in\mathbb R_+,$$
then the~$f(r)$-modified N--P charges associated to~$\psi$ can be read off from Table~\ref{tab:conformalregularity} below:
\begin{table}[htpb]
\centering
\begin{tabular}{l|lll}
   Values of $p_\ell,q_\ell$  & {($p_\ell< \ell,q_\ell>0$) or~$(p_\ell=\ell, q_\ell>0)$:} & {$(p_\ell=\ell, q_\ell=0)$:}            &{($p_\ell>\ell,q_\ell\geq 0$):}                  \\ \hline
{$\pv\Phi_\ell=$}& {$\frac{A\log^{q_\ell} r}{r^{2-(\ell-p_\ell)}}+\dots$} & {$\frac{B}{r^2}+\dots$} &{$\frac{B}{r^2}+\frac{A\log^{q_\ell} r}{r^{2-(\ell-p_\ell)}}+\dots$} 
\end{tabular}
\caption{The letter~$A$ is a placeholder for a nonvanishing constant multiple of~$C^*$, whereas~$B$ stands for a linear combination of the~$C_i$ and may therefore be vanishing. The second row of the table indicates for which $f$ the~$f(r)$-modified N--P charge is finite and nonvanishing. For instance, if $p_\ell<\ell$, then we can take $f(r)=r^{\ell-p_\ell-2}\log^{q_\ell} r$ and $I_\ell^f[\psi]=A$. On the other hand, if the data satisfy peeling, then we always have $f(r)=r^{-2}$.} 
\label{tab:conformalregularity}
\end{table}

%
%

\paragraph{The~$f(r)$-modified N--P charges for higher spin~$s$:}
In a similar fashion, one can define conserved charges for more general spin. 
For instance, if~$s=\pm2$ and $M=0$, then the generalisation of the Minkowskian identities \eqref{eq:NP:NPMink} is given by
\begin{equation}\label{eq:NP:Teukolsky}
\pu(r^{-2N-2s}\pv(r^2\pv)^{N}(r^{|s|+s+1}\Psi^{[|s|-s]}_{\ell}))=\frac{(N+s-\ell)(N+s+\ell+1)}{r^{2N+2s+2}}(r^2\pv)^{N}(r^{|s|+s+1}\Psi^{[|s|-s]}_{\ell}),
\end{equation}
where~$\Psi^{[|s|-s]}_{\ell}$ is the projection of~$\Psi^{[|s|-s]}$ onto the spin-$s$ weighted spherical harmonics~${}_sY_{\ell m}$, which are defined for~$\ell\geq |s|$ (here, we again suppressed the~$m$-index).
From \eqref{eq:NP:Teukolsky}, one can then derive an equation similar to \eqref{eq:NP:NPl} in order to derive the relevant~$f(r)$-modified N--P charges for~$M\neq 0, |s|=2$.
Comparing \eqref{eq:NP:NPMink} with \eqref{eq:NP:Teukolsky}, one thus finds that, roughly speaking, \textbf{the~$\ell$-th mode of~$r^{|s|+s}\Psi^{[|s|-s]}$ behaves like the~$(\ell-s)$-mode of~$\psi$ would}, as the RHS of \eqref{eq:NP:Teukolsky} vanishes for $N=\ell-s$.


\section[Deriving asymptotics towards $i^+$ from asymptotics towards~$\mathcal I^+$]{From asymptotics towards~$\mathcal I^+$ to asymptotics towards~$i^+$}\label{sec:goodnighttales}
We will now sketch the derivation of the leading-order late-time asymptotics in time~$\tau$ as~$\tau\to \infty$ of solutions to the wave equation \eqref{eq:wave} \textbf{starting from initial data on the hyperboloidal initial hypersurface}~$\Sigma_0$. The arguments in this section generalise arguments from~\cite{AAG18a, AAG18b, AAG21,II}. 

We first consider data for which there exists $f_\ell$ such that $I_{\ell}^{f_\ell}[\psi]\neq 0$ in~\S\ref{sec:nonzeroNP}. As we will see, the late-time tails for $\psi_\ell$ are directly encoded in the value of $I_{\ell}^{f_\ell}[\psi]\neq 0$ in this case.
In~\S\ref{sec:zeroNP}, we then consider data for which $I_{\ell}^{f_\ell}[\psi]=0$ for any choice of $f_\ell$, and reduce this case to that of~\S\ref{sec:nonzeroNP}.
The analyses of~\S\ref{sec:nonzeroNP} and~\S\ref{sec:zeroNP} produce the late-time tails for fixed angular frequency solutions $\psi_\ell$. We comment on general solutions in \S\ref{sec:summing}.

\subsection{The case of nonvanishing N--P charge~$I_{\ell}^f[\psi]$}
\label{sec:nonzeroNP}
We assume smooth initial data on $\Sigma_0$ and take~$f_{\ell}$ to be a function of~$r$ with the following general form:
\begin{equation}\label{eq:LT:f}
f_{\ell}(r)=r^{-p'_{\ell}}(\log r)^{q_{\ell}},
\end{equation}
where~$1-\ell<p'_{\ell}<3$ and~$q_{\ell}\geq 0$, $p_\ell'\in\mathbb R$, $q_\ell\in\mathbb N$. 
Given $f_\ell$, we make the following assumptions on the~$r$-asymptotics of the initial data on~$\Sigma_0$: The $\ell$-th spherical harmonic mode~$\psi_{\ell}$ satisfies
\begin{align}
\label{eq:iddecay1}
\begin{split}
\partial_v \Phi_{\ell}|_{\Sigma_0}=&\:I_{\ell}^{f_{\ell}}[\psi] f_{\ell}(r)+O(r^{-p'_{\ell}})\quad \textnormal{if~$q_{\ell}>0$},\\
\partial_v \Phi_{\ell}|_{\Sigma_0}=&\:I_{\ell}^{f_{\ell}}[\psi] f_{\ell}(r)+O(r^{-p'_{\ell}-\beta})\quad \textnormal{if~$q_{\ell}=0$},
\end{split}
\end{align}
with~$I_{\ell}^{f_{\ell}}[\psi]\neq 0$,~$\beta>0$, and with~$\Phi_{\ell}$ defined in \eqref{PHIL}. 

In Steps 0--3 below, we outline how we can \underline{translate} the above initial data~$r$-asymptotics to the following late-time~$\tau$-asymptotics and~$u$-asymptotics:
\begin{align}
\label{eq:tails1}
\psi_{ \ell}|_{r=r_0}(\tau)=&\: A_{\ell}w_\ell(r_0)I_{\ell}^{f_{\ell}}[\psi] f_{\ell}(\tau) \tau^{-2\ell}+\ldots\quad (\tau\to \infty),\\
\label{eq:tails2}
r\psi_{\ell}|_{\mathcal{I}^+}(u)=&\:A_{\ell} I_{\ell}^{f_{\ell}}[\psi] f_{\ell}(u) u^{1-\ell}+\ldots\quad (u\to \infty),
\end{align}
with~$A_{\ell}\in \mathbb{R}$ constants that depend only on~$p,q,\ell$, and~$w_\ell(r_0)$ also depending on~$r_0$ (see \eqref{eq:definitionofw} for the precise~$r_0$-dependence), and where~$\ldots$  schematically denote terms that contribute as higher-order terms in~$\tau^{-1}$ or~$u^{-1}$.

\paragraph{Step 0:}
The mechanism for deriving late-time tails relies on the following type of \emph{upper bound time-decay estimate}:
\begin{equation}
\label{eq:almostsharp}
|r^{-\ell}\psi_{\ell}|\leq B_{\epsilon} (\tau+1)^{1-p'_{\ell}-\ell+\epsilon}(\tau+r)^{-\ell-1},\\
\end{equation}
with~$\epsilon>0$ arbitrarily small and~$B_{\epsilon}>0$ an appropriately large constant depending on~$L^2$-type initial data norms and diverging as $\epsilon\to 0$. 
In light of the expected time-decay that can be read off from \eqref{eq:tails1} and \eqref{eq:tails2}, the estimate \eqref{eq:almostsharp} can be thought of as an \emph{almost sharp} time-decay estimate.

We will moreover make use of the fact that, in $(\tau,r)$-coordinates, when acting with the vector fields~$\partial_{\tau}$,~$r\partial_r$ and~$r^2\partial_r$ on~$r\psi_{\ell}$, the~$\tau$-decay rate in \eqref{eq:almostsharp} changes according to Table \ref{tbl:tabledtdr}:
\begin{table}[ht]
\caption{Change in decay rate when acting with weighted vector fields on~$\psi_{\ell}$.}
\centering
\begin{tabular}{c|c}
Vector field & Change in power of~$\tau$-factor in  \eqref{eq:almostsharp}\\
\hline
$\partial_{\tau}$ & -1 \\
$r\partial_r$ & +0 \\ 
$r^2\partial_r$ & +1
\label{tbl:tabledtdr}
\end{tabular}
\end{table}

The estimate \eqref{eq:almostsharp} and the properties of Table \ref{tbl:tabledtdr} are slight generalisations of what is derived in~\cite{AAG18a, AAG18b, AAG21}.  
The methods used to derive \eqref{eq:almostsharp} build on the vast literature on uniform boundedness and decay estimates for linear waves on black hole spacetime backgrounds and involve geometric properties of Schwarzschild like the trapping of null geodesics and the red-shift effect; see~\cite{DRSR16} for a comprehensive overview of this literature.
A derivation of these almost-sharp estimates lies beyond the scope of the present paper, so we will view them simply as \emph{black box assumptions} in a self-contained derivation of late-time tails.

In Steps 1--3 below, we will outline the derivation of the precise leading-order late-time asymptotics and late-time tails for the spherically symmetric~$\ell=0$-mode. We will briefly describe the generalization to~$\ell\geq 1$ afterwards.
\paragraph{Step 1:}
Let $\alpha\in(0,1)$. We first restrict to the region~$\{v\geq v_{\gamma}(u)\}$. 
Here,~$\gamma$ is a timelike curve along which $$r= u^{\alpha}+\ldots=v^{\alpha}+\ldots,$$with~$\ldots$ denoting terms that are higher order in~$u^{-1}$ and~$v^{-1}$, and $v_\gamma(u)$ denotes the unique $v$ such that $(u,v)\in\gamma$. 
In this region, we apply \eqref{eq:NP:NPl=0} and \eqref{eq:iddecay1} to obtain:
\begin{align*}
\partial_u \partial_v\phi_0=O(r^{-3})\phi_0+\ldots,\quad \textnormal{with }
\partial_v\phi_0|_{\Sigma_0}(v)= I_0^{f_0}[\psi] f_0\left(\tfrac{v}{2}\right)+\ldots,
\end{align*}
with~$\ldots$ again denoting terms that contribute to sub-leading order in the argument. Here, we used moreover that~$r=\frac{1}{2}(v-u)+\ldots$ to leading order in~$v-u$, so~$r|_{\Sigma_0}=\frac{v}{2}+\ldots$.
\begin{figure}[htpb]
\includegraphics[width=130pt]{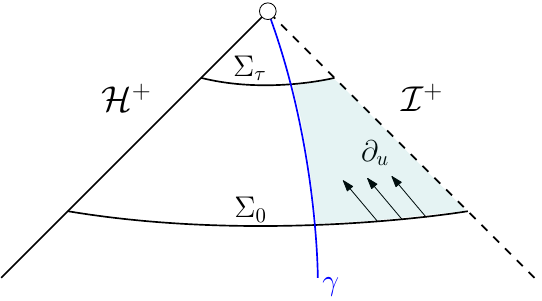}
\caption{Step 1: integrating~$\partial_u\partial_v\phi_0$ in~$u$.}
\label{fig:uint}
\end{figure}

Since~$\alpha>0$,~$r$ \emph{grows} along~$\gamma$, and the factor~$r^{-3}$ on the right-hand side of~$\partial_u \partial_v\phi_0$ \emph{decays} in the region~$\{v\geq v_{\gamma}(u)\}$. 
Therefore, we can integrate the equation for~$\partial_u\partial_v\phi_0$ in the~$u$-direction (see Fig.~\ref{fig:uint}) and plug in the almost-sharp estimate \eqref{eq:almostsharp} on the right-hand side to \underline{propagate} the asymptotics of~$\partial_v\phi_{0}$ from~$\Sigma_0$ to the rest of the region~$\{v\geq v_{\gamma}(u)\}$:
\begin{align}
\label{eq:step1}
\partial_v\phi_{0}(u,v)=&\:  I_{0}^{f_0}[\psi]f_0\left(\tfrac{v}{2}\right)+\ldots.
\end{align}
Note that \eqref{eq:step1} in particular implies  the following modified N--P conservation law along~$\mathcal{I}^+$: The limit~$\lim_{r\to \infty}\frac{1}{f_0(r)} \partial_v\phi_0(u,r)$ is conserved in~$u$ along~$\mathcal{I}^+$ and is equal to~$I_{0}^{f_0}[\psi]$.

\paragraph{Step 2:}

\begin{figure}[htpb]
\includegraphics[width=130pt]{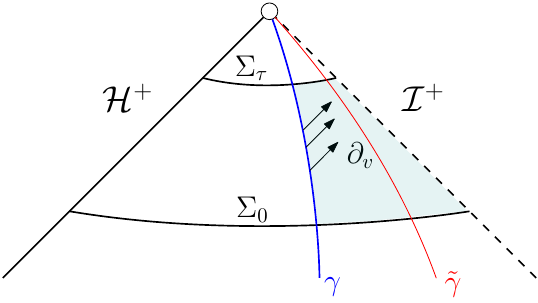}
\caption{Step 2: integrating~$\partial_v\phi_{0}$ in~$v$ from~$\gamma$.}
\label{fig:vint}
\end{figure}

In this step, we integrate \eqref{eq:step1} in the~$v$-direction  in the region~$\{v\geq v_{\gamma}(u)\}$, now starting from~$\gamma$ (see Fig.~\ref{fig:vint}):
\begin{equation*}
\phi_{0}(u,v)=\phi_{0}|_{\gamma}(u)+\int_{v_{\gamma}(u)}^v\partial_v\phi_{0}(u,v')\dd v'.
\end{equation*}
The curve $\gamma$ is chosen such that~$\phi_0|_{\gamma}$ only contributes to higher order. Indeed, using our definition of~$\gamma$ ($r_\gamma\sim u^{\alpha}$ along~$\gamma$), we can split
\begin{equation*}
|\phi_{0}|_{\gamma}|=r_{\gamma} |\psi_0|_{\gamma}|\sim u^{\alpha} |\psi_0|_{\gamma}|.
\end{equation*}
Since~$\alpha<1$, and since $u\sim\tau$ along $\gamma$, we can then apply the estimate \eqref{eq:almostsharp} with~$\epsilon>0$ suitably small to conclude that~$\phi_{0}|_{\gamma}$ in fact decays  \emph{faster} than~$u^{-p_0'+1}$. Hence,
\begin{equation*}
\begin{split}
\phi_{0}(u,v)=\int_{v_{\gamma}(u)}^v\partial_v\phi_{0}(u,v')\dd v'+\ldots=I_{0}^{f_0}[\psi]\int_{v_{\gamma}(u)}^v f_0(\tfrac{v'}{2}) \dd v'+\ldots.
\end{split}
\end{equation*}
Given the form of~$f_0$ assumed in \eqref{eq:LT:f}, we can evaluate the above integral to obtain:
\begin{equation*}
\begin{split}
\phi_0(u,v)=&\:I_{0}^{f_0}[\psi]\int_{v_{\gamma}(u)}^v(\tfrac{v'}{2})^{-p_0'} \log^{q_0} (\tfrac{v'}{2})\dd v'+\ldots\\
=&\:\tfrac{2^{p_0'}}{p_0'-1} I_{0}^{f_0}[\psi]\cdot \left(\log^{q_0}( \tfrac{v_{\gamma}(u)}{2})(v_{\gamma}(u)^{-p_0'+1}-v^{-p_0'+1})+v^{-p_0'+1}\log^{q_0}(\tfrac{v_{\gamma}(u)}{v})\right)\ldots
\end{split}
\end{equation*}
everywhere in~$\{v\geq v_{\gamma}(u)\}$, where \ldots\ here denote terms that will contribute as higher order terms in~$u^{-1}$ or~$\tau^{-1}$ below. In particular, restricting to a smaller region~$\{v\geq v_{\tilde{\gamma}}(u)\}$, where~$\tilde{\gamma}$ is a timelike curve along which~$r=u^{\tilde{\alpha}}+\ldots= v^{\tilde{\alpha}}+\ldots$, for~$\alpha<\tilde{\alpha}<1$ with~$\tilde{\alpha}$ suitably close to~1, we obtain:
\begin{equation}
\label{eq:asympl01}
\phi_{0}(u,v)=\tfrac{2^{p_0'}}{p-1}I_{0}^{f_0}[\psi](\log^{q_0}( u) (u^{-p_0'+1}-v^{-p_0'+1})+\ldots.
\end{equation}
Taking the limit~$v\to\infty$, we thus obtain
\begin{align*}
\phi_0|_{\mathcal{I}^+}(u)=&\:\tfrac{2^{p_0'}}{p_0'-1}I_{0}^{f_0}[\psi]u^{-p_0'+1}\log^{q_0} u+\ldots,
\end{align*}
which proves \eqref{eq:tails2} for~$\ell=0$.
Moreover, by Taylor expanding the expression~$\frac{2}{v_{\tilde{\gamma}}(u)-u}(u^{-p+1}-v_{\tilde{\gamma}}(u)^{-p+1})$  in~$\frac{v_{\tilde{\gamma}}(u)-u}{u}=u^{\tilde{\alpha}-1}$ around 0, we can also obtain the late-time asymptotics of~$\psi_0$ along~$\tilde \gamma$:
\begin{align}\label{eq:LT:gammatilde}
\psi_0|_{\tilde{\gamma}}(u)=\tfrac{2\phi_0|_{\tilde{\gamma}}(u)}{v_{\tilde{\gamma}}(u)-u}+\ldots=&\: 2^{p_0'+1} I_{0}^{f_0}[\psi]u^{-p_0'}\log^{q_0} u +\ldots.
\end{align}
Notice, in particular, that the $\log^{q_0} u$-term in \eqref{eq:LT:gammatilde} directly corresponds to the $\log^{q_0} r$-term in~\eqref{eq:LT:f}.

\paragraph{Step 3:}
\begin{figure}[htpb]
\includegraphics[width=180pt]{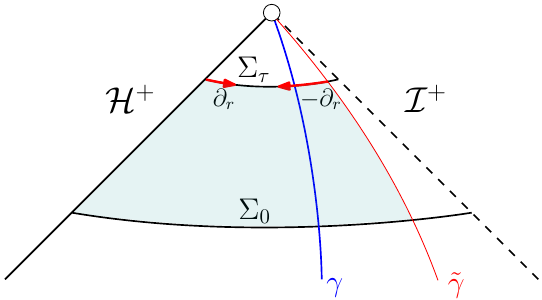}
\caption{Step 3: integrating~$\partial_r(Dr^{2}\partial_r\psi_0)$ along~$\Sigma_{\tau}$ from~$r=2M$ and then~$\partial_r \psi_{0}$ from~$\tilde{\gamma}$.}
\label{fig:twowayint}
\end{figure}
To also conclude \eqref{eq:tails1}, we \emph{propagate} the late-time asymptotics \eqref{eq:LT:gammatilde} derived along the curve~$\tilde{\gamma}$ in Step 2 all the way to the event horizon at~$r=2M$. We use that, in~$(\tau,r,\theta,\varphi)$ coordinates, the wave equation for~$\psi_{0}$ takes the following schematic form:
\begin{equation}
\label{eq:weightedwe}
\partial_r((r^2-2Mr)\partial_r\psi_0)=O(r)\cdot \left(\partial_{\tau}\psi_{0}+ r\partial_r\partial_{\tau}\psi_{0}+\partial_{\tau}^2\psi_{0}\right).
\end{equation}
We integrate eq.~\eqref{eq:weightedwe} from~$r=2M$ (see Fig.~\ref{fig:twowayint}) to obtain everywhere in~$2M\leq r\leq r_{\tilde{\gamma}}(\tau)$:
\begin{equation*}
(r^2-2Mr)|\partial_r\psi_0|(\tau,r)\lesssim \int_{2M}^r r\left(|\partial_{\tau}\psi_{0}|+ |r\partial_r\partial_{\tau}\psi_{0}|+|\partial_{\tau}^2\psi_{0}|\right)\dd r'.
\end{equation*}

Now, we can exploit the fact that the right-hand side above involves terms with at least one~$\partial_{\tau}$-derivative: 
Indeed, the estimate \eqref{eq:almostsharp} together with Table \ref{tbl:tabledtdr} tells us that the integrand on the RHS goes like $r\cdot \tau^{-p_0'-1+\epsilon}$. 
Doing the integral, we thus obtain, for all~$r\leq r_{\tilde{\gamma}}(\tau)$:
\begin{equation}
\label{eq:asympnonzeroNPremain}
|\partial_r\psi_{0}|\leq C(1+\tau)^{-p_0'-1+\epsilon}.
\end{equation}

Finally, for any $r\in[2M,r_{\tilde{\gamma}}(\tau)]$, 
 we write $\psi_0(\tau,r)=\psi_0(\tau,r_{\tilde{\gamma}}(\tau))+\int \partial_r\psi_0(\tau,r')\dd r'$, and combine estimates \eqref{eq:LT:gammatilde} and \eqref{eq:asympnonzeroNPremain}.
Again, the integral term only contributes at higher order, so we obtain
\begin{equation}
\label{eq:asympl02}
\psi_{0}(\tau,r)=\psi_{0}(\tau,r_{\tilde{\gamma}}(\tau))+\ldots=2^{p_0'+1} I_{0}^{f_0}[\psi]\cdot \tau^{-p_0'}(\log \tau)^{q_0} +\ldots
\end{equation}
everywhere in~$2M\leq r\leq r_{\tilde{\gamma}}(\tau)$. This concludes the derivation of \eqref{eq:tails1} in the case~$\ell=0$.

\paragraph{Treatment of~$\ell\geq 1$ modes:}

The argument for higher~$\ell$-modes runs similarly, with the role of~$\phi_0$ now being played by~$\Phi_\ell$. We start \textbf{Step 1} by taking \eqref{eq:NP:NPl}, which schematically reads
\begin{align*}
\partial_u(D^{\ell} r^{-2\ell}\partial_v\Phi_{\ell})=&\:O(r^{-3-2\ell})\Phi_{\ell}+\ldots,
\end{align*}
inserting the estimates \eqref{eq:almostsharp} into the RHS, and finally integrating in $u$. 
This proves the analogue of \eqref{eq:step1}, giving us a sharp estimate for $\pv\Phi_\ell$ and the conservation of~$I_\ell^{f_\ell}[\psi]$.
Then, in \textbf{Step 2,} we integrate~$\partial_v\Phi_{\ell}$ in~$v$ to first obtain the late-time behaviour of~$\Phi_{\ell}$, and we subsequently use that
\begin{equation*}
\Phi_{\ell}(u,v)=(D^{-1}r^2\partial_v)^{\ell}(r\psi_{\ell})+\ldots,
\end{equation*}
with~$\ldots$ terms that contribute as higher-order terms, to obtain the late-time behaviour of~$\phi_{\ell}$ after another~$\ell$ integrations in~$v$. 
Since each of these integrations picks up an extra~$1/u$-term from a suitably chosen $\gamma$, this thus proves \eqref{eq:tails2}, as well as the analogue of \eqref{eq:LT:gammatilde}.

Finally, in \textbf{Step 3}, we consider the weighted quantity~$w_{\ell}^{-1}(r)\psi_{\ell}$, with~$w_{\ell}$ a smooth, non-zero radial weight function that satisfies
\begin{align}\label{eq:definitionofw}
\square_g(w_{\ell} Y_{\ell m})=0\quad \textnormal{for all~$|m|\leq \ell$},\quad \textnormal{and}\quad
w_{\ell}(r)\sim &\:r^{\ell}\quad  \textnormal{as~$r\to \infty$}.
\end{align}
The product~$w_{\ell}^{-1}(r)\psi_{\ell}$ then satisfies the equation:
\begin{equation}
\label{eq:weightedweell}
\partial_r((r^2-2M r)w_{\ell}^2\partial_r(w_{\ell}^{-1}\psi_{\ell}))=O(r^{\ell+1})\cdot\left(\partial_{\tau}\psi_{\ell}+ r\partial_r\partial_{\tau}\psi_{\ell}+\partial_{\tau}^2\psi_{\ell}\right),
\end{equation}
which involves only terms with~$\partial_{\tau}$-derivatives on the right-hand side. We integrate this equation to show that~$\partial_r(w_{\ell}^{-1}\psi_{\ell})$ is higher-order in~$\tau^{-1}$. One more integration in~$r$ then results in the global late-time asymptotics for~$w_{\ell}^{-1}\psi_{\ell}$ and, in particular, proves \eqref{eq:tails1}.

\subsection{Vanishing~$I_{\ell}^f[\psi]$}
\label{sec:zeroNP}
When~$I_{\ell}^{f_{\ell}}[\psi]=0$ in the initial data assumptions \eqref{eq:iddecay1}, the steps outlined in Section~\ref{sec:nonzeroNP} cannot be applied directly. 
We now explain how to proceed in this case.
Let~$f_{\ell}(r)=r^{-p'_{\ell}}(\log r)^{q_{\ell}}$ with~$p'_{\ell}\geq 2$, and assume the following $r$-asymptotics on initial data:
\begin{align}
\label{eq:iddecayNP01}
\partial_v \Phi_{\ell}|_{\Sigma_0}=&\: J_{\ell}r^{-1}f_{\ell}(r)+O(r^{-p'_{\ell}-1})\quad \textnormal{if~$q_{\ell}>0$},\\
\label{eq:iddecayNP02}
\partial_v \Phi_{\ell}|_{\Sigma_0}=& J_{\ell} r^{-1}f_{\ell}(r)+O(r^{-p'_{\ell}-1-\beta})\quad \textnormal{if~$q_{\ell}=0$},
\end{align}
for some~$J_{\ell}\in\mathbb R$. 
Assumption \eqref{eq:iddecayNP02} implies that~$I_{\ell}^{f}[\psi]=0$ for any choice of~$f$. 
Furthermore, since \underline{we allow~$J_{\ell}$ to be zero}, these assumptions include initial data of compact support.

The object that is key to deriving late-time tails for these rapidly~$r$-decaying initial data is the \emph{time integral}:
\begin{equation}
\partial_{\tau}^{-1}\psi_{\ell}(\tau,r):=-\int_{\tau}^{\infty} \psi_{\ell}(\tau',r)\dd \tau'.
\end{equation}
By the decay estimates \eqref{eq:almostsharp} in Step 0,~$\partial_{\tau}^{-1}\psi_{\ell}$ is well-defined and regular at the event horizon. 
Moreover, by definition,~$\partial_{\tau}(\partial_{\tau}^{-1}\psi_{\ell})=\psi_{\ell}$, and, by the symmetries of the Schwarzschild spacetime, $\partial_{\tau}^{-1}\psi_{\ell}$ still solves \eqref{eq:wave} projected to $\ell$.
Lastly, we can determine the value of~$\partial_{\tau}^{-1}\psi_{\ell}|_{\Sigma_0}$ via simple integration of:
\begin{equation}
\partial_r((r^2-2M r)w_{\ell}^{2}\partial_r(w_{\ell}^{-1}\partial_{\tau}^{-1}\psi_{\ell}))|_{\Sigma_0}=O(r^{\ell+1})\left[\psi_{\ell}+ r\partial_r\psi_{\ell}+\partial_{\tau}\psi_{\ell}\right]|_{\Sigma_0},
\end{equation}
using that~$(r^2-2Mr)w_{\ell}^2\partial_r(w_{\ell}^{-1}\partial_{\tau}^{-1}\psi_{\ell})$ vanishes at~$r=2M$ and~$w_{\ell}^{-1}\partial_{\tau}^{-1}\psi_{\ell}$ vanishes as~$r\to \infty$. 

It then follows that the asymptotics of~$\partial_v \partial_{\tau}^{-1}\Phi_{\ell}$ can be grouped into two cases:
\begin{align*}
\partial_v \partial_{\tau}^{-1}\Phi_{\ell}|_{\Sigma_0}=& (I_{\ell}^{(1)})^{f_{\ell}}[\psi] f_{\ell}(r)+O(r^{-p'_{\ell}})\quad \textnormal{if~$q_{\ell}>0$ \underline{and}~$p'_{\ell}=2$ \underline{and}~$J_{\ell}\neq 0$},\\
\partial_v  \partial_{\tau}^{-1} \Phi_{\ell}|_{\Sigma_0}=&(I_{\ell}^{(1)})^{r^{-2}}[\psi] r^{-2}+O(r^{-2-\beta'})\quad \textnormal{with~$\beta'>0$, if~$q_{\ell}=0$ \underline{or}~$p'_{\ell}>2$ \underline{or}~$J_{\ell}=0$},
\end{align*}
where~$(I_{\ell}^{(1)})^f[\psi]$ is the~$f(r)$-modified N--P charge of~$\partial_{\tau}^{-1}\psi$, which can be expressed either as a constant multiple of~$J_{\ell}$ if~$q_{\ell}>0$,~$p'_{\ell}=2$ and~$J_{\ell}\neq 0$, or in terms of an integral of initial data for~$\psi$ along~$\Sigma_0$ otherwise.
In the latter case,~$(I_{\ell}^{(1)})^{r^{-2}}[\psi]$ is \emph{generically nonvanishing if~$M\neq 0$}.\footnote{Here, ``generic'' can be given a precise meaning:~$(I_{\ell}^{(1)})^f[\psi]=0$ only for a codimension-1 subset of data in any suitably~$r$-weighted function space. In fact, within this codimension-1 subset, we can consider \emph{another} time integral, so we apply time integration twice. Using multiple time inversions, we can conclude that the set of data leading to solutions that do \underline{not} behave inverse polynomially in time is therefore of infinite codimension.}

We can now apply the arguments of \S\ref{sec:nonzeroNP} to obtain the late-time asymptotics of~$\partial_{\tau}^{-1}\psi_{\ell}$ (cf.~\eqref{eq:tails1}, \eqref{eq:tails2}) and finally take a~$\partial_{\tau}$-derivative to deduce the asymptotics for~$\psi$ itself.
 We obtain
\begin{align}
\label{eq:tails1timeint}
\psi_{ \ell}|_{r=r_0}(\tau,\theta,\varphi)=&\: A_\ell w_{\ell}(r_0)(I_{\ell}^{(1)})^f[\psi] \frac{d}{d\tau}(f(\tau) \tau^{-2\ell})+\ldots\quad(\tau\to \infty),\\
\label{eq:tails2timeint}
r\psi_{ \ell}|_{\mathcal{I}^+}(u,\theta,\varphi)=&\:A_{\ell} (I_{\ell}^{(1)})^f[\psi] \frac{d}{du}(f(u) u^{1-\ell})+\ldots\quad (u\to \infty),
\end{align}
with~$f(x)$ either given by~$f_\ell(x)$ or by~$x^{-2}$.
Thus, for rapidly decaying initial data, \textbf{the~$\tau$-decay rates in the corresponding late-time tails are encoded in the~$r$-decay of the initial data of the\textit{ time integral}~$\partial_{\tau}^{-1}\psi$!}

The results of \S \ref{sec:nonzeroNP} and \S \ref{sec:zeroNP} are summarised in the table below, where~$f_\ell=r^{-p_\ell'}\log^{q_\ell} r$ with~$p'_{\ell}>1-\ell$ and~$q_{\ell}\geq 0$:
\begin{table}[htpb]
\centering
\begin{tabular}{l|ll}
 &$p_{\ell}'<3$ or~$p_{\ell}'=3$ and~$q_{\ell}>0$:    &~$p_{\ell }'> 3$  or~$p_{\ell}'=3$ and~$q_{\ell}=0$:  \\ \hline
 $\pv\Phi_\ell|_{\Sigma_0}$ as~$r\to\infty$:    &~$I_{\ell}^{f_\ell}[\psi]\cdot f_\ell(r)+\dots$  &~$J_{\ell}\cdot f_{\ell}(r)+\ldots$              \\ 
 $\phi_\ell|_{\mathcal I^+}$ as~$u\to\infty$:&			$A_\ell I_{\ell}^{f_\ell}[\psi] \cdot f_{\ell}(u) u^{1-\ell}+\dots$		&		$A_\ell   \cdot (I_{\ell}^{(1)})^{r^{-2}}[\psi] u^{-2-\ell}+\dots$ 	\\
$\psi_\ell|_{r=r_0}$ as~$\tau\to\infty$:&		$A_{\ell}  w_{\ell}(r_0)I_{\ell}^{f_\ell}[\psi]\cdot f_\ell(\tau)\tau^{-2\ell}+\dots$	 &	$A_{\ell}   w_{\ell}(r_0)\cdot(I_{\ell}^{(1)})^{r^{-2}}[\psi]\cdot \tau^{-3-2\ell}+\dots$ 			
\end{tabular}
\caption{Summary of the results of \S\ref{sec:goodnighttales}. The~$A_\ell$ are placeholders for nonvanishing numerical constants. The function~$w(r_0)$ is defined in \eqref{eq:definitionofw}. 
The rates in the second column are independent of the mass~$M$, whereas the constant~$(I_{\ell}^{(1)})^{r^{-2}}[\psi]$ appearing in the third column does depend on~$M$.}
\label{tab:summary latetimes}
\end{table}

\subsection{Summing over~$\ell$}\label{sec:summing}
In \S \ref{sec:nonzeroNP} and \S \ref{sec:zeroNP}, we considered the late-time asymptotics of spherical harmonic modes of fixed~$\ell$. 
When choosing initial data such that higher~$\ell$-modes have more regularity in~$\frac{1}{r}$ as~$r\to \infty$, and, in particular, when considering smooth, compactly supported data, higher modes will decay faster. 
In such cases, the estimate \eqref{eq:almostsharp} can be extended to
\begin{equation}\label{eq:almostsharpsummed}
|r^{-\ell}\psi_{\geq \ell}|\leq C \tau^{1-p'_{\ell}-\ell+\epsilon}(\tau+r)^{-\ell-1}.
\end{equation}
Then, by \eqref{eq:almostsharpsummed}, it follows that when splitting~$\psi_{\geq \ell}=\psi_{\ell}+\psi_{\geq \ell+1}$, the~$\psi_{\geq \ell+1}$ contributes at higher order in~$\tau^{-1}$ and~$u^{-1}$, so the late-time asymptotics of~$\psi_{\geq \ell}$ agree with the late-time asymptotics of~$\psi_{ \ell}$, which are given by \eqref{eq:tails1} and \eqref{eq:tails2}, or \eqref{eq:tails1timeint} and \eqref{eq:tails2timeint}. 
However, initial data on~$\Sigma_0$ for which higher modes do \underline{not} have higher regularity at infinity may arise naturally in scattering problems, see \S 1.3.3  of~\cite{III}. 
There, it is conjectured that compactly supported scattering data lead to solutions where \textit{all modes} contribute to the late-time asymptotics at the same order. 
See already the fourth row, second column of Table~\ref{tab:ET:full}.

\section{Deriving asymptotics towards~$\mathcal I^+$ from physical data near~$i^-$}\label{sec:goodmorningtales}
In \S\ref{sec:goodnighttales}, we derived late-time asymptotics near~$i^+$ from given asymptotics along some hyperboloidal initial data hypersurface~$\Sigma_0$, i.e.\ from asymptotics \textit{towards~$\mathcal I^+$}.
Naturally, to decide what the "correct" predictions for late-time asymptotics are, one thus needs a way to decide what the  "correct"  asymptotics towards~$\mathcal I^+$ are. 
As explained in the introduction, instead of \textit{assuming}, say, vanishing or "peeling" asymptotics near~$\mathcal I^+$, we will \textit{dynamically derive} the asymptotics towards~$\mathcal I^+$ from a scattering data setup that
\textbf{a)}
 has no incoming radiation from~$\mathcal I^-$ and
\textbf{b)} resembles---in some sense---a system of~$N$ infalling masses following unbound Keplerian orbits near the infinite past. 
This section follows the works~\cite{I,II,III}.
\subsection{The data setup}
\begin{figure}[htpb]
\includegraphics[width=120pt]{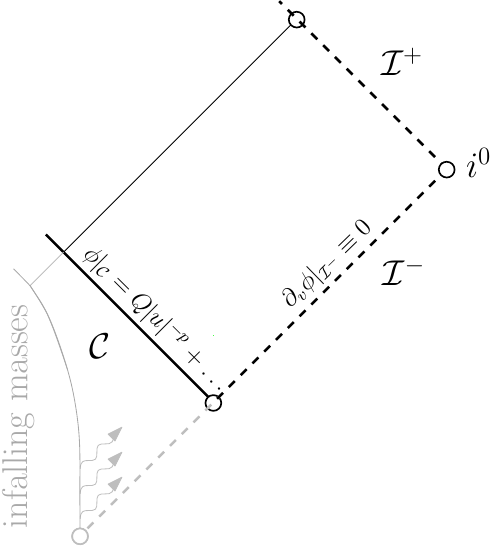}
\caption{Depiction of the data setup: We aim to capture the radiation of the infalling masses by imposing the quadrupole approximation on~$\mathcal C$. The depicted spacetime can be thought of as part of the spacetime depicted in Fig.~1.}\label{fig:ET}
\end{figure}

In the context of the scalar wave equation on a fixed Schwarzschild background, one simple model with data that realise \textbf{a)} and \textbf{b)} is depicted in Fig.~\ref{fig:ET}:
In order to satisfy \textbf{a)} on~$\mathcal I^-$, we demand~$\pv(r\psi)|_{\mathcal I^-}\equiv0$ to be vanishing identically. 
This corresponds to a vanishing energy flux along~$\mathcal I^-$. 
Realising~\textbf{b)}, on the other hand, is less straightforward. 
The idea is as follows: While it may, for now, be too ambitious to try and analytically treat a system of~$N$ infalling masses, we can instead, for sufficiently large negative retarded times~$u$, consider an ingoing null cone~$\mathcal C$ from~$\mathcal I^-$, to be thought of as enclosing the~$N$ infalling masses, and impose data on~$\mathcal C$ that capture the structure of the radiation emitted by these~$N$ infalling masses, see Fig.~\ref{fig:ET}.
The \textit{heuristic tool} that allows us to take this step is the \textit{quadrupole approximation}, which predicts that (gravitational) radiation decays polynomially along~$\mathcal C$ with a certain rate.

Identifying this radiation with scalar radiation $\psi$, we now give a brief sketch of where this polynomial decay comes from:
The quadrupole approximation for~$N$ infalling masses predicts that the loss of gravitational energy along~$\mathcal I^+$ is given by
$$\frac{dE_\text{grav}}{dt}\sim-\dddot{\mathbf{Q}}_{ij}^{TT}\dddot{\mathbf{Q}}_{ij}^{TT},$$
the $\mathbf{Q}_{ij}$ denoting the quadrupole moment of the mass distribution. 
In the case of hyperbolic orbits\footnote{\label{footnotepara}We could do a similar analysis for parabolic orbits, which would give the exponent $10/3$ instead of 4 in \eqref{exp4}. } (i.e.\ if the relative velocities of the masses tend to nonzero constants {in the infinite past}), one thus gets that (see \cite{Chr02} for a derivation)
\begin{equation}\label{exp4}
\frac{dE_\text{grav}}{dt}\sim-C|u|^{-4}+\dots \quad \textnormal{as~$u\to-\infty$}.
\end{equation}
We can now identify this gravitational energy along $\mathcal I^+$ with the scalar field energy, namely the flux of the Noether current associated to~$T=\partial_t$:
\begin{equation}
\frac{dE_\text{grav}}{dt}\sim\frac{dE_{\text{scalar}}}{dt}=-\int_{\mathbb S^2}(\pu\phi|_{\mathcal I^+})^2 {\dd\sigma_{\mathbb{S}^2}}.
\end{equation}
In this sense, the quadrupole approximation "predicts" that~$|\pu\phi|_{\mathcal I^+}|\sim |u|^{-2}$, and hence~$|\phi|_{\mathcal I^+}|\sim |u|^{-1}$. 

One can make a  more sophisticated argument at the level of gravitational (i.e.\ not scalar) perturbations ($s=\pm2$) that also {allows} one {to obtain} such a rate on {an} ingoing null hypersurface at a \textit{finite}~$v$-distance~$\mathcal C$ as opposed to $\mathcal I^+$~\cite{WalkerWill79,Damour86}.
This justifies the assumption that~$\phi|_{\mathcal C}=Q|u|^{-p}+\dots$ for~$p=1$, $Q\neq 0$.

In fact, a more detailed perturbative analysis would result in different predictions on the exponent~$p$ for each angular mode~$\ell$.
For the purpose of this expository note, we will simply make the assumption that
\begin{equation}
\phi_\ell|_{\mathcal C}=Q_\ell |u|^{-p_\ell}+\dots
\end{equation}
for some general~$p_{\ell}\in\mathbb R_+$,~$Q_\ell\in\mathbb R\setminus\{ 0\}$, keeping in mind that for~$\ell=0$,~$p_0=1$ is the "physically relevant" exponent.\footnote
{Of course, we can only speak of physical relevance in the case of gravitational radiation, i.e.\ for~$s=\pm 2$, which we discuss in \S\ref{sec:ET:grav}. 
However, since the quadrupole approximation gives a prediction for the lowest angular modes of~$\Psi^{[0]}$ and~$\Psi^{[4]}$, namely the~$\ell=2$ modes, and the lowest angular mode of~$\psi$ is given by~$\ell=0$, we here take~$p_0=1$ to be physically relevant exponent for the scalar field. 
See also the arguments relating the analysis of the $s=0$ case to the $s=\pm 2$ case in \S\ref{sec:dictionary} (cf.~Footnote~\ref{footnote}).}

We will now complete our dictionary by explaining how these asymptotics along~$\mathcal C$ translate into asymptotics towards~$\mathcal I^+$, first for~$\ell=0$ in \S\ref{sec:ET:l=0}, then for general~$\ell$ in \S\ref{sec:ET:l}. 
We will give an outlook on the case of~$s=\pm 2$ in \S\ref{sec:ET:grav}.

\subsection{Analysis of the spherically symmetric part~$\phi_0$}\label{sec:ET:l=0}
We now explain how to treat solutions~$\psi_0$ arising from a data setup as in Fig.~\ref{fig:ET}, i.e.\ with data~$\phi_0|_{\mathcal C}(u)=Q_0|u|^{-p_0}+\dots$ on $\mathcal C$ and~$\pv\phi_0|_{\mathcal I^-}\equiv 0$ on~$\mathcal I^-$.
These data are to be understood as \textit{scattering data}. 
Recently developed scattering theory~\cite{nicolas,DRSR18,Masaood22} ensures the unique existence of a solution~$\psi_0$ that attains the prescribed data along~$\mathcal C$ and~$\mathcal I^-$, and moreover provides us with the (far from optimal) global bound that (see also \cite{I})
\begin{equation}
|r\psi_0|\leq C\sqrt{r}.
\end{equation}

To obtain asymptotic estimates, the analysis will roughly follow two steps:
\begin{itemize}
\item[ \textbf{(I)}] We first insert the weak estimate~$|\phi_0|\lesssim \sqrt{r}$ into \eqref{eq:NP:NPl=0} and integrate \eqref{eq:NP:NPl=0} from~$\mathcal I^-$ to obtain a weak estimate on~$\pv\phi_0$. 
\item[ \textbf{(II)}] We can then integrate this estimate for~$\pv\phi_0$ from~$\mathcal C$ to obtain an estimate for~$\phi_0$.
\end{itemize}
The crucial observation then is that, as a consequence of the strong~$r^{-3}$-weight appearing on the right-hand side of \eqref{eq:NP:NPl=0}, this new estimate on~$\phi_0$ will be improved compared to the original one. 
One can then iterate steps \textbf{(I)} and \textbf{(II)} until one obtains a sharp estimate.
Now, the details:

\textbf{(I):} Inserting~$|\phi_0|\leq \sqrt r$ into \eqref{eq:NP:NPl=0}, and integrating \eqref{eq:NP:NPl=0} from~$\mathcal I^-$, we find
\begin{equation}\label{eq:ET:step1:l=0}
\pv\phi_0(u,v)=\int_{-\infty}^u \pu\pv\phi_0\dd u'\leq \int_{-\infty}^u \frac{MCD}{2r^{5/2}}\dd u'\leq C r^{-3/2}.
\end{equation}
Here, we used the no-incoming-radiation condition~$\pv\phi_0|_{\mathcal I^-}\equiv 0$. 

\textbf{(II):} Integrating \eqref{eq:ET:step1:l=0} from~$\mathcal C$ (where~$2r|_{\mathcal C}\sim |u|$), we then find
\begin{equation}\label{eq:ET:step2:l=0}
|\phi_0(u,v)-\phi|_{\mathcal C}(u)|\leq C \int_{1}^v r^{-3/2}\dd v'\leq C |u|^{-1/2}.
\end{equation}
This clearly improves the initial estimate~$|\phi_0|\leq C\sqrt{r}$ to~$|\phi_0|\leq C|u|^{-\min(p_0,1/2)}$.

One can now insert this improved estimate back into \eqref{eq:ET:step1:l=0} and iteratively repeat the procedure of steps \textbf{(I)} and \textbf{(II)}, i.e.\ of~\eqref{eq:ET:step1:l=0}, \eqref{eq:ET:step2:l=0}, to eventually obtain the estimate
\begin{equation}\label{eq:ET:asy1:l=0}
|\phi_0(u,v)-\phi_0|_{\mathcal C}(u)|\leq C |u|^{-p_0-1}.
\end{equation}
Since~$\phi_0|_{\mathcal C}(u)=Q_0 |u|^{-p_0}+\dots$, \eqref{eq:ET:asy1:l=0} {implies that the leading-order term in the asymptotics of~$\phi_0$ as~$u\to-\infty$ is $Q_0 |u|^{-p_0}$.}

\begin{figure}[htpb]
\includegraphics[width=120pt]{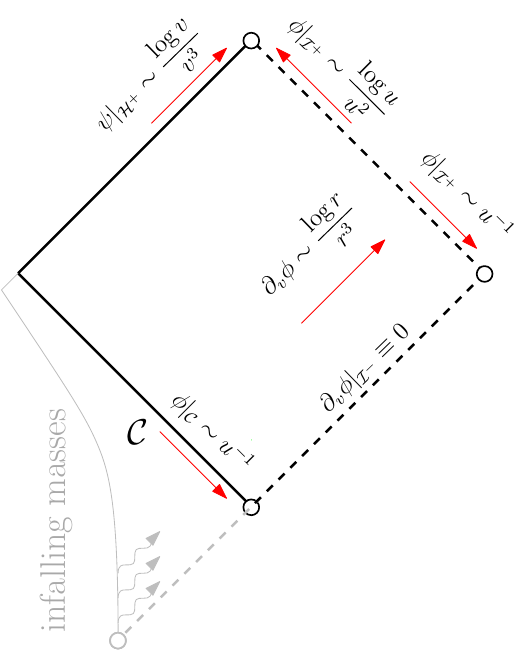}
\caption{Depiction of the resulting asymptotics in various regions if~$\phi|_{\mathcal C}\sim u^{-1}$ initially. }\label{fig:ETtheorem}
\end{figure}

Finally, in order to also obtain an asymptotic estimate for~$\pv\phi_0$ as~$v\to\infty$, we once again repeat the calculation \eqref{eq:ET:step1:l=0}, this time equipped with the asymptotic estimate \eqref{eq:ET:asy1:l=0}.
This gives
\begin{equation}\label{eq:ET:asy2:l=0}
\pv\phi_0(u,v)=\int_{-\infty}^u \pu\pv\phi_0\dd u'= \int_{-\infty}^u -\frac{MDQ_0|u|^{-p_0}}{2r^3}\dd u'+\dots.
\end{equation}
The integral on the RHS can be computed by writing~$2r=v-u+\mathcal O(\log(v-u))$.
Let us here give the concrete computation only in the  case~$p_0=1$:
\begin{multline}
\int_{-\infty}^u\frac{1}{(v-u')^3u'}\dd u'= \int_{-\infty}^u \frac{1}{v^3}\left(\frac{1}{u' }+\frac{1}{ v-u'}+\frac{v}{ (v-u')^2}+\frac{v^2}{(v-u')^3}\right)\dd u'
\\=       \frac{\log|u|-\log(v-u)}{v^3}+\frac{3v-2u}{2v^2(v-u)^2}.
\end{multline}
In particular, we conclude that, along~$u=const$ (where~$v\sim r$), we have 
\begin{equation}\label{eq:57}
\pv\phi_0(u,v)=-\frac{MQ_0}{2}\frac{\log r-\log |u|}{r^3}+\mathcal O(r^{-3}).
\end{equation}
Similar computations can be done for general~$p_0$, see Table~\ref{tab:ET:l0} below:
\begin{table}[htpb]
\centering
\begin{tabular}{l|lllll}
$\phi_0|_{\mathcal C}=Q_0|u|^{-p_0}+\dots$                        &$p_0\in(0,1)$:               &$p_0=1$:                       &$p_0\in(1,2)$:                               &$p_0=2$:                                        &$\dots$ \\ \hline
$\pv\phi_0(u=const,v)=$&$\frac{AM}{r^{2+p_0}}+$ &$\frac{AM\log r}{r^3}+$ &$\frac{h(u)}{r^3}+\frac{AM}{r^{2+p_0}}+$ &$\frac{h(u)}{r^3}+\frac{AM\log r}{r^4}+$ &$\dots$ 
\end{tabular}
\caption{Relation between decay of~$\phi_0$ as~$u\to-\infty$ and asymptotics for~$\pv\phi_0$ as~$v\to\infty$. Here,~$h(u)$ is a function decaying in~$u$ as~$u\to-\infty$, and~$A$ stands for a constant multiple of~$Q_0$.}
\label{tab:ET:l0}
\end{table}\\
In particular, if~$p_0=1$, then \eqref{eq:57} implies that~$I_0^{f_0}[\psi]=-MQ_0/2$ for~$f_0=r^{-3}\log r$. Thus, combining the findings of the present section and of \S \ref{sec:goodnighttales} (cf.~Table~\ref{tab:summary latetimes}), we can {in particular} conclude the following~\cite{I,II}:
\begin{thm}\label{thm}
Consider data on~$\mathcal C$ and~$\mathcal I^-$ that satisfy~$\phi_0|_{\mathcal C}=Q_0|u|^{-1}+\mathcal O(|u|^{-1-\epsilon})$ as~$u\to-\infty$ and~$\pv\phi_0|_{\mathcal I^-}\equiv0$. 
Then, along any outgoing null hypersurface of constant~$u$,~$2\pv\phi_0=-MQ_0r^{-3}\log r+\dots$ as~$v\to\infty$, so~$I_0^{{r^{-3}\log r}}[\psi]\equiv -MQ_0/2$. 

Moreover, if one smoothly extends the data along~$\mathcal C$ to~$\mathcal H^+$ as in Fig.~\ref{fig:ETtheorem}, one obtains the following late-time asymptotics near~$i^+$:\footnote{Note that, owing to the conservation of~$I_0^{r^{-3}\log r}[\phi]$ along~$\mathcal I^+$, the leading-order asymptotics \eqref{5.11} are  independent of the extension of the data towards $\mathcal H^+$. }
\begin{align}\label{5.11}
\psi_0|_{\mathcal H^+}=-4MQ_0v^{-3}\log v+\dots &&\phi_0|_{\mathcal I^+}=-2MQ_0u^{-2}\log u+\dots.
\end{align} 

\end{thm}
Finally, we point out that, regardless of the value of $p_0$, $\phi_0$ will never fully satisfy peeling towards $\mathcal I^+$ under this setup, as indicated by Table~\ref{tab:ET:l0}.
\subsection{Higher~$\ell$-modes~$\psi_\ell$}\label{sec:ET:l}
We now turn our attention to higher~$\ell$-modes: We consider data such that~$r\psi_\ell|_{\mathcal C}(u)=Q_\ell|u|^{-p_\ell}+\dots$ on~$\mathcal C$ and such that~$\pv(r\psi_\ell)|_{\mathcal I^- }\equiv 0$ along~$\mathcal I^-$.

The analysis of higher~$\ell$-modes will again be similar to the analysis of the~$\ell=0$ mode, with the relevant equation now being \eqref{eq:NP:NPl} instead of \eqref{eq:NP:NPl=0}.
For simplicity, let us work with a simplified version of\ \eqref{eq:NP:NPl}, namely (recall~$\VV :=D^{-1}r^2\pv$):
\begin{equation}\label{eq:NP:simplified}
\pu(D^\ell r^{-2\ell}\pv \VV ^\ell\phi_\ell)=\frac{MD^{\ell+1}y_0^{(\ell)}}{r^{2\ell+3}} \VV ^\ell\phi_\ell.
\end{equation}
This simplification corresponds to setting~$x_i^{(\ell)}=0=y_i^{(\ell)}$ for all~$i>0$ and~$z_i^{(\ell)}=0$ for all~$i\geq 0$ in \eqref{eq:NP:NPl}, but still allows us to capture the main ideas of the proof: 
Essentially, we can now repeat a procedure almost identical to steps \textbf{(I)} and \textbf{(II)} of \S\ref{sec:ET:l=0}, with~$\phi_0$ replaced by~$\Phi_\ell:=\VV ^\ell\phi_\ell$.
The main difference is that, in order to get an estimate similar to \eqref{eq:ET:asy1:l=0}, one first needs to compute the values of~$\Phi_\ell$ along~$\mathcal C$.
This is achieved by inductively integrating the equations satisfied by~$\pu(r^{-2N}\pv \VV ^N\phi_\ell)$ for~$N<\ell$ from~$u=-\infty$, with these equations in turn being obtained by simply commuting the wave equation \eqref{eq:NP:wave} with~$\VV ^N$ (cf.\ \eqref{eq:NP:NPMink}).
This gives:
\begin{equation}\label{eq:ET:dataterms}
\VV ^N\phi_\ell|_{\mathcal C}(u)=A_N^{(\ell)} |u|^{-p_\ell+N}+\dots
\end{equation}
for some constants~$A_N^{(\ell)}$ that are nonvanishing multiples of~$Q_\ell$ for~$N\leq \ell$.

Equipped with this estimate for~$\Phi_\ell|_{\mathcal C}$, we can now, similarly to how we showed \eqref{eq:ET:asy1:l=0}, show that
\begin{equation}\label{eq:ET:L:asystep1}
|\Phi_\ell(u,v)-\Phi_\ell|_{\mathcal C}(u)|\leq C \begin{cases}
r^{\ell-1-p_\ell}& \text{ if } p_\ell<\ell -1\\
\log r-\log|u| &\text{ if } p_\ell=\ell-1\\
|u|^{\ell-1-p_\ell}& \text{ if } p_\ell>\ell-1
\end{cases}
\end{equation} 
(Note that the first two cases in \eqref{eq:ET:L:asystep1} did not appear for~$\ell=0$ since we assumed~$p_\ell$ to be positive. Formally, however, the calculations here are valid for \textit{any} value of~$p_\ell\in \mathbb R$.)
As with the~$\ell=0$-case, we can now insert this estimate 
into \eqref{eq:NP:simplified} to obtain an asymptotic estimate on~$\pv\Phi_\ell$:
\begin{equation}\label{eq:ET:L:asystep2}
\pv\Phi_\ell(u,v)
=r^{2\ell}\int_{-\infty}^u  \frac{MDy_0^{(\ell)}}{r^{2\ell+3}}\Phi_\ell|_{\mathcal C}\dd u'+\ldots
=\begin{cases} 
A r^{\ell-p_\ell-2}+\dots & \text{ if } p_\ell< \ell+1\\
A r^{-3}(\log r-\log|u|)+\dots & \text{ if } p_\ell= \ell+1\\
A |u|^{\ell+1-p_\ell} r^{-3} +\dots & \text{ if } p_\ell> \ell+1
\end{cases}
\end{equation}
Notice that if $p_\ell\leq \ell+1$, then the RHS of \eqref{eq:ET:L:asystep2} gives the relevant $f(r)$-modified N--P charge and shows that it is conserved as well. For instance, for $p_\ell<\ell+1$ and $f_\ell=r^{\ell-p_\ell-2}$, we have~$I_{\ell}^{f_\ell}[\psi]=A$. On the other hand, if $p_\ell>\ell+1$, then all $f(r)$-modified N--P charges vanish.

Finally, by integrating \eqref{eq:ET:L:asystep2}~$\ell+1$ times from~$\mathcal C$ (where $2r=|u|+\ldots$), each time picking up a term on~$\mathcal C$ that is given by \eqref{eq:ET:dataterms}, we can now also obtain an estimate for~$\phi_\ell$ itself.\footnote{
We resort to an example in order to schematically explain this.
Consider the~$\ell=1$-mode~$\psi_1$ with~$p_1>0$.
From the estimate \eqref{eq:ET:L:asystep1}, we obtain that~$r^2\pv\phi_1(u,v)=r^2\pv\phi_1|_{\mathcal C}(u)+\dots$.
We further compute from \eqref{eq:NP:wave} that~$r^2\pv\phi_1|_{\mathcal C}(u)=A_1^{(1)}\cdot |u|^{-p_1+1}+\dots$, with~$A_1^{(1)}=-Q_1/(2(p_1+1))$ (cf.~\eqref{eq:ET:dataterms}).
Therefore, we can  schematically compute~$\phi_1(u,v)$ via
\begin{equation*}
\phi_1(u,v)=\phi_1|_{\mathcal C}(u)+\int \pv\phi_1(u,v)\dd v=\frac{Q_1}{|u|^{p_1}}+\frac{A_1^{(1)}}{|u|^{p_1-1}}\int r^{-2}\dd v+\ldots=\frac{Q_1}{|u|^{p_1}}-\frac {Q_1}{(p+1)|u|^{p_1-1}}\left(\frac{1}{r|_{\mathcal C}}-\frac1r\right)+\ldots.
\end{equation*}
In particular,~$\lim_{v\to\infty}\phi_1(u,v)$ decays faster in~$u$ than~$\phi_1|_{\mathcal C}(u)$ does initially iff~$p_1=1$.

More generally, if~$\phi_\ell|_{\mathcal C}$ decays like~$|u|^{-p_\ell}$ initially, then~$\phi_\ell|_{\mathcal{I}^+}$ will decay faster than~$|u|^{-p_\ell}$ if and only if~$p_\ell\in\{1,\dots,\ell\}$.} 
A more detailed discussion of this can be found in~\cite{III}.

%
\section{Completing the dictionary and extending to gravitational perturbations}\label{sec:dictionary}
\paragraph{Completing the dictionary:}
We can finally summarise our findings and combine the results of \S \ref{sec:goodmorningtales} with those of \S\ref{sec:goodnighttales}.
Suppose that we have data for~$\psi_\ell$ such that~$\phi_\ell|_{\mathcal C}=Q_\ell|u|^{-p_\ell}+\dots$ near~$\mathcal I^-$ and such that~$\pv\phi_\ell|_{\mathcal I^-}\equiv 0$.
We can then read off the relevant choice of~$f(r)$-modified N--P charge from \eqref{eq:ET:L:asystep2}.
Thus, combining this with the results of \S \ref{sec:goodnighttales}, cf.~Table~\ref{tab:summary latetimes} (and again smoothly but arbitrarily extending the data towards~$\mathcal H^+$), we can directly connect the behaviour of~$\phi_\ell$ near~$\mathcal I^-$ to its behaviour  near~$i^+$.
This is done in Table~\ref{tab:ET:full} below.
\begin{table}[htpb]
\centering
\begin{tabular}{l|lll}
            $\phi_\ell|_{\mathcal C} =Q_\ell |u|^{-p_\ell}+\dots$ as~$u\to-\infty$:             			   &~$p_\ell<\ell+1$:     &~$p_\ell=\ell+1$:           &~$p_\ell>\ell+1$: \\ \hline
$\phi_\ell|_{\mathcal I^+}$ as~$u\to-\infty$:&\multicolumn{3}{c}{$A|u|^{-p_\ell}$ unless~$p_\ell\in\{1,\dots,\ell\}$}\\
$\pv \Phi_\ell$ as~$u=const, v\to\infty$:   &$AM r^{-2+\ell-p_\ell}$ &$AMr^{-3}\log r$ &$Br^{-3}$         \\ 
$\phi_\ell|_{\mathcal I^+}$ as~$u\to\infty$:&			$AM u^{-1-p_\ell}$		&		$AM u^{-2-\ell}\log u$		&$B u^{-2-\ell}$ \\
$\psi_\ell|_{\mathcal H^+}$ as~$v\to\infty$:&		$ AM v^{-2-\ell-p_\ell}$			 &	$ AM v^{-2\ell-3}\log v$			&$ B v^{-2\ell-3}$
\end{tabular}
\caption{The full thesaurus: The letter~$A$ is always a placeholder for a constant, nonvanishing multiple of~$Q_\ell$ (that is different from cell to cell), whereas~$B$ stands for a constant that also depends on the extension of the data along~$\mathcal C$ towards~$\mathcal H^+$ and is only generically nonvanishing.}
\label{tab:ET:full}
\end{table}
\paragraph{Extending to gravitational perturbations  ($s=\pm 2$):}\label{sec:ET:grav}

So far, we have focussed mostly on~$s=0$. 
The more realistic case of gravitational perturbations will be discussed in detail in upcoming work~\cite{IV,V}, but we shall already list the main points here:
Post-Newtonian arguments~\cite{WalkerWill79,Damour86} predict the following rates along~$\mathcal C$ at the level of quadrupolar radiation (i.e.\ for~$\ell=2$):~$\Psi^{[0]}\sim |u|^{-3}$ and~$\Psi^{[4]}\sim |u|^{-4}$ as~$u\to-\infty$.\footnote{The reader could have loosely guessed these rates using the same simplistic heuristics as we presented for~$s=0$: The rate~$r\Psi^{[4]}|_{\mathcal C}\sim |u|^{-3}$ as $u\to-\infty$ on data will remain the same on~$\mathcal I^+$, where~$r\Psi^{[4]}=\pu N$ gives the rate of change of the News function $N$. 
Furthermore, we have the Bondi mass loss formula~\cite{SeriesVII,SeriesVIII,CK93} along $\mathcal I^+$:~$\frac{dE_\text{grav}}{dt}=-\frac{1}{4\pi}\int_{\mathbb S^2}|N|^2{\dd\sigma_{\mathbb{S}^2}}$. 
The rate for $r\Psi^{[4]}$ thus comes from the quadrupole approximation prediction that~$\frac{dE_\text{grav}}{dt}\sim -|u|^{-4}$. 
The rate for~$\Psi^{[0]}$ is then enforced by the Teukolsky--Starobinsky identities~\cite{Teukolsky74} and the no-incoming-radiation condition. \label{footnote}} 
Combining this with the no-incoming-radiation condition, a preliminary analysis then gives:
\begin{enumerate}
\item[$\upalpha)$] In addition to~$\Psi^{[4]}$ violating the peeling rate near~$\mathcal I^-$ (which states that~$\Psi^{[4]}=\mathcal O(r^{-5})$), the peeling rate of~$\Psi^{[0]}$ near~$\mathcal I^+$ is also violated: Instead of the peeling rate~$\Psi^{[0]}=\mathcal O(r^{-5})$, one will obtain that~$\Psi^{[0]}\sim r^{-4}$ towards~$\mathcal I^+$. 
Motivations for this rate have appeared before in~\cite{Damour86,Chr02}.
\item[$\upbeta)$] While the radiation field for~$\Psi^{[0]}$,~$\lim_{\mathcal I^+}r^5\Psi^{[0]}$, thus blows up, the radiation field for~$\Psi^{[4]}$, $\lim_{\mathcal I^+}r\Psi^{[4]}$, is still defined, and we conjecture that the failure of peeling (i.e.\ of conformal regularity) in $\upalpha)$ translates into the following late-time decay rate along~$\mathcal I^+$:
$r\Psi^{[4]}|_{\mathcal I^+}\sim u^{-3}$ as~$u\to\infty$. 
This should be contrasted with the Price's law rate, which predicts that~$r\Psi^{[4]}|_{\mathcal I^+}\sim u^{-6}$; see~\cite{MZ21} for a derivation of the Price's law rate for $|s|=2$. 
\end{enumerate} 

In fact, at a heuristic level, the reader can already guess the rates in $\upalpha)$ and $\upbeta)$ by the observation from \S\ref{sec:NP} that the following correspondences hold:
\begin{align*}
r^5\Psi^{[0]}_{\ell=2} \leftrightarrow  \phi_{\ell=0},&&
r\Psi^{[4]}_{\ell=2}\leftrightarrow  \phi_{\ell=4}.
\end{align*}
Therefore, our rates on~$\mathcal C$ read:
\begin{align*}
|u|^2\sim r^5\Psi^{[0]}_{\ell=2}|_{\mathcal{C}}\leftrightarrow &\: \phi_{\ell=0}|_{\mathcal{C}}\quad \implies p_0=-2,\\
|u|^{-3}\sim r\Psi^{[4]}_{\ell=2}|_{\mathcal{C}}\leftrightarrow &\: \phi_{\ell=4}|_{\mathcal{C}}\quad \implies p'_4=3.
\end{align*}
\textit{However, the behaviour of $\Psi^{[4]}_{\ell=2}$ is actually governed by $p_4=2$, not $p'_4=3$.} Leaving the details to \cite{IV,V}, we here only note that this is related to the fact that, as a consequence of the extra $r^{-2s}$-weight in \eqref{eq:NP:Teukolsky}, the $|u|^{-3}$-decay of $r\Psi^{[4]}_{\ell=2}|_{\mathcal{C}}$ leads to a non-integrable RHS of \eqref{eq:NP:Teukolsky} for $N=0$ when trying to compute the transversal derivatives of $r\Psi^{[4]}_{\ell=2}$ along $\mathcal C$ as in \eqref{eq:ET:dataterms}: This leads to $\VV (r\Psi^{[4]})$ decaying only like $|u|^{-2}\log|u|$ along $\mathcal{C}$, and to $\VV^2(r\Psi^{[4]})\sim 1$ not decaying along $\mathcal C$. So, compared to \eqref{eq:ET:dataterms}, higher-order transversal derivatives decay one power slower compared to the case $s=0$.

By now applying Table~\ref{tab:ET:full} with these correspondences and values for~$p_{\ell}$, we obtain in particular:
\begin{align}
\partial_v(r^5\Psi^{[0]}_{\ell=2})|_{u={ const}}\leftrightarrow  \partial_v\phi_{\ell=0}|_{u={ const}}\overset{p_0=-2}\sim r^{-2+0-(-2)}=r^0 \quad &\text{as }r\to \infty\nonumber \\
\overset{\textnormal{integrating}}\implies \Psi^{[0]}_{\ell=2}|_{u={ const}}\sim r^{-4} \quad &\text{as }r\to \infty,
\\
 r\Psi^{[4]}_{\ell=2}|_{\mathcal{I}^+}\leftrightarrow  \phi_{\ell=4}|_{\mathcal{I}^+}\overset{p_4=2}\sim u^{-1-2}=u^{-3}\quad &\text{as }u\to \infty.
\end{align}

The numerology presented above is rooted in the assumption of hyperbolic Keplerian orbits in the infinite past and \eqref{exp4}. Using similar arguments, the reader can also find the numerology in the case of \textit{parabolic orbits}, cf.~Footnote \ref{footnotepara}.
%

\section{Further directions}\label{sec:Extending}
We append the main body of the paper with some remarks on extensions of the presented methods.

\paragraph{From Schwarzschild to subextremal Kerr:} For solutions to the scalar wave equation on subextremal Kerr arising from conformally regular or compactly supported initial data, the leading-order late-time asymptotics feature the same rates as in Schwarzschild~\cite{Hintz22,AAG23}. 
See also~\cite{SRTdC20,SRTdC23, MZ23} for related recent results in the setting of the Teukolsky equations and \cite{barack99grav, LBAO99} for a heuristic analysis of late-time tails in Kerr spacetimes. 

The effects of the non-zero angular momentum of Kerr do, however, affect the decay rates of higher angular modes. 
Since Kerr is not spherically symmetric, there is no \emph{a priori} canonical definition of spherical harmonics and corresponding angular modes,\footnote{Note however that in phase space, the Boyer--Lindquist time-frequency-dependent spheroidal harmonics form a natural choice of angular modes, since they are involved in the separability of the wave equation after taking a Fourier transform in time.} and, whichever definition is chosen, obtaining late-time tails for each mode will involve the difficulty of \emph{mode coupling}. 

This difficulty was addressed and studied in~\cite{AAG23}, where it was shown that the choice of spherical harmonics with respect to Boyer--Lindquist spheres at infinity allows for a modified analogue of Price's law in Kerr. The topic of generalising Price's law on Kerr spacetimes has a long history featuring various conflicting predictions. See \cite{zengi14,burko14} and references therein for an overview of this problem and for the latest numerical results, which are in alignment with the mathematically rigorous results derived in \cite{AAG23}.

We expect that the techniques outlined in the present paper will also be applicable in combination with the methods developed in~\cite{AAG23} to study the effects that a violation of peeling has on late-time tails for the Teukolsky equations on subextremal Kerr backgrounds.

\paragraph{Extremal black holes:} 
\textit{Extremal} black holes feature several additional fascinating phenomena that have an effect on the rates of decay in late-time tails and are inherently connected to the degeneracy of extremal event horizons. 
For instance, the wave equation on extremal Reissner--Nordstr\"om black holes and the axisymmetric wave equation on extremal Kerr black holes possess additional conserved charges along the event horizon~\cite{Aretakis15} that lead to different decay rates from the subextremal setting~\cite{AAG20} and are connected to the presence of asymptotic instabilities known in the literature as the \emph{Aretakis instabilities}~\cite{Aretakis15}; see also earlier heuristics in \cite{ori2013, sela}. 
In fact, in extremal Reissner--Nordstr\"om, these additional conservation laws can be related to the Newman--Penrose charges by applying the Couch--Torrence conformal isometry that maps null infinity to the event horizon~\cite{CouchTorrence84, Bizo__2013,Lucietti2013}. 

In upcoming work (see~\cite{Gajic21Oberwolfach,Gajic23Kerr}), it is shown that the late-time tails of \underline{non}-axisymmetric solutions to the wave equation on extremal Kerr exhibit even stronger deviations from the subextremal case, as well as stronger instabilities, consistent with the heuristics in~\cite{glampedakisfull,zimmerman1}. 
Due to the lack of conserved charges along the event horizon in the non-axisymmetric setting, we moreover need a different mechanism for deriving late-time tails from the one presented in the present paper.\footnote{A similar lack of conserved charges occurs also in the model problems of scalar fields with an inverse-square potential on Schwarzschild, and a new mechanism is developed in~\cite{Gajic22Inverse} to overcome this.} 

Late-time tails in gravitational radiation are of particular observational relevance in the extremal setting since they have been predicted to form the dominant part of gravitational wave signals at much earlier stages in the ringdown process than in the subextremal setting~\cite{zengi13}. 
They therefore provide a promising observational signature of (near)-extremality of black holes. 
Since extremal late-time tails also decay much slower, and since they are a phenomenon associated to regularity at the future event horizon and not at future null infinity, \textit{we do not expect the failure of peeling to have an effect on decay rates,} in contrast to the subextremal setting.

\paragraph{Moving beyond linear perturbations:}
The results of \S\ref{sec:goodmorningtales} have been extended to the coupled Einstein-scalar field system under the assumption of spherical symmetry in~\cite{I}. 
The effects of nonlinearities on Price's law (i.e.\ including backreaction) have been investigated heuristically and numerically in~\cite{Bizon09, Bizon10}, where deviations to Price's law have been predicted for higher spherical harmonic modes arising from compactly supported data. 
See also the recently announced results of Luk--Oh on late-time tails on fixed, but \emph{dynamical} black hole spacetime backgrounds, where mathematically rigorous methods are applied to obtain similar deviations \cite{Luktalk21}. 
The above works suggest that in the full nonlinear theory, compactly supported Cauchy data should lead to the following late-time tails along~$\mathcal{I}^+$:
\begin{equation*}
r\Psi^{[4]}|_{\mathcal{I}^+}\sim u^{-5}.
\end{equation*}
While this is slower than the~$u^{-6}$-tail expected from linear theory (for compactly supported data), this late-time tail still decays \emph{faster} than the~$u^{-3}$-tail that we predicted above as a consequence of the failure of peeling. 
In light of this, we expect that, even in the fully nonlinear theory, the dominant late-time behaviour will be~$u^{-3}$, provided that we consider the physically motivated scattering data of the present paper.

Finally, in view of the impressive techniques that have been developed to study the dynamics of gravitational radiation in the setting of the full system of the \textit{nonlinear Einstein vacuum equations}~\cite{CK93,DHRT21}, we expect a mathematically rigorous investigation of the above prediction to be within reach.

\section{Conclusion}\label{sec:conclusions}
We want to conclude with the following points:
\begin{itemize}
\item[$\diamond$] If one has initial data on some hyperboloidal hypersurface~$\Sigma_0$ for which one can define nonvanishing~$f(r)$-modified N--P charges, then the late-time asymptotics towards~$i^+$ can be read off from these N--P charges according to Table~\ref{tab:summary latetimes}.
The crucial question then is: What is the right choice of~$f(r)$-modified N--P charge? 
\item[$\diamond$]
If one poses polynomially decaying data on some ingoing null hypersurface emanating from~$\mathcal I^-$ and excludes radiation coming in from~$\mathcal I^-$, then, because of the nonvanishing background mass~$M$ near spatial infinity~$i^0$, the backscatter of gravitational radiation at early times will lead to~$\mathcal I^+$ not being smooth, and this failure of smoothness will determine the choice of~$f(r)$-modified N--P charge and, therefore, the late-time tails of gravitational radiation near~$i^+$, see Table~\ref{tab:ET:full}.
The smoothness of~$\mathcal I^-$ plays no role here.
\item[$\diamond$] The assumption of polynomial decay towards~$\mathcal I^-$, in turn, comes from post-Newtonian arguments and the assumption that the system under consideration, e.g.\ two infalling masses, follows approximately hyperbolic Keplerian orbits in the infinite past. 
\end{itemize}
We record again that it is frequently assumed throughout large parts of the literature that one has spatially compact support on~$\Sigma_0$, or that gravitational radiation has only started radiating at some fixed, finite time (which of course implies the former).
However, in the context of an isolated system describing an astrophysical process, we believe the assumptions of the present paper, i.e.\ that the system under consideration has radiated \textit{for all times}, to be more natural.

Independently of the above considerations, we also hope to have convinced the reader that, even from a purely theoretical point of view, the assumption of smooth null infinity might be too rigid, and that by avoiding this assumption, one can perform many more general arguments that give new and deeper insights into the nature of general relativity!
\small{\bibliographystyle{alpha} 
\bibliography{references_all}}

\newcommand{\etalchar}[1]{$^{#1}$}
\begin{thebibliography}{BVdBM62}

\bibitem[A{\etalchar{+}}16]{ligotest}
B.~P. Abbott et~al.
\newblock {Tests of general relativity with GW150914}.
\newblock {\em Phys. Rev. Lett.}, 116(22):221101, 2016.
\newblock [Erratum: Phys.Rev.Lett. 121, 129902 (2018)].

\bibitem[A{\etalchar{+}}21]{LIGOlatetime}
R.~Abbott et~al.
\newblock {Tests of General Relativity with GWTC-3}.
\newblock {\em arXiv:2112.06861}, 2021.
\newblock [Accepted in Phys. Rev. D].

\bibitem[AAG18a]{AAGLetters}
Y.~Angelopoulos, S.~Aretakis, and D.~Gajic.
\newblock Horizon hair of extremal black holes and measurements at null
  infinity.
\newblock {\em Phys. Rev. Lett.}, 121(13):131102, 2018.

\bibitem[AAG18b]{AAG18b}
Y.~Angelopoulos, S.~Aretakis, and D.~Gajic.
\newblock Late-time asymptotics for the wave equation on spherically symmetric,
  stationary backgrounds.
\newblock {\em Adv. in Math.}, 323:529--621, 2018.

\bibitem[AAG18c]{AAG18a}
Y.~Angelopoulos, S.~Aretakis, and D.~Gajic.
\newblock A vector field approach to almost-sharp decay for the wave equation
  on spherically symmetric, stationary spacetimes.
\newblock {\em Annals of PDE}, 4(2), 2018.

\bibitem[AAG20]{AAG20}
Y.~Angelopoulos, S.~Aretakis, and D.~Gajic.
\newblock Late-time asymptotics for the wave equation on extremal
  {R}eissner--{N}ordstr\"{o}m backgrounds.
\newblock {\em Adv. in Math.}, 375, 2020.

\bibitem[AAG21]{AAG21}
Y.~Angelopoulos, S.~Aretakis, and D.~Gajic.
\newblock {Price's law and precise asymptotics for subextremal
  Reissner--Nordstr\"om black holes}.
\newblock {\em arXiv:2102.11888}, 2021.

\bibitem[AAG23]{AAG23}
Yannis Angelopoulos, Stefanos Aretakis, and Dejan Gajic.
\newblock Late-time tails and mode coupling of linear waves on kerr spacetimes.
\newblock {\em Adv. in Math.}, 417, 2023.

\bibitem[Are15]{Aretakis15}
S.~Aretakis.
\newblock Horizon instability of extremal black holes.
\newblock {\em Adv. Theor. Math. Phys.}, 19:507--530, 2015.

\bibitem[BCR09]{Bizon09}
P.~Bizo{\'n}, T.~Chmaj, and A.~Rostworowski.
\newblock {Late-time tails of a self-gravitating massless scalar field,
  revisited}.
\newblock {\em Class. Quantum Grav.}, 26(17):175006, 2009.

\bibitem[BF13]{Bizo__2013}
Piotr Bizo{\'{n}} and Helmut Friedrich.
\newblock {A remark about wave equations on the extreme
  Reissner{\textendash}Nordström black hole exterior}.
\newblock {\em Class. and Quantum Grav.}, 30:065001(6), February 2013.

\bibitem[BK14]{burko14}
L.~M. Burko and G.~Khanna.
\newblock {Mode coupling mechanism for late-time Kerr tails}.
\newblock {\em Phys. Rev. D}, 89(4):044037, 2014.

\bibitem[BKS19]{burko19}
L.~M. Burko, G.~Khanna, and S.~Sabharwal.
\newblock Transient scalar hair for nearly extreme black holes.
\newblock {\em Phys. Rev. Res.}, 1(3):033106, 2019.

\bibitem[BO99a]{barack99grav}
L.~Barack and A.~Ori.
\newblock {Late-time decay of gravitational and electromagnetic perturbations
  along the event horizon}.
\newblock {\em Phys. Rev. D}, 60(12):124005, 1999.

\bibitem[BO99b]{LBAO99}
L.~Barack and A.~Ori.
\newblock Late-time decay of scalar perturbations outside rotating black holes.
\newblock {\em Phys. Rev. Lett.}, 82(4388-4391), 1999.

\bibitem[BR10]{Bizon10}
P.~Bizo{\'n} and A.~Rostworowski.
\newblock {Note about late-time wave tails on a dynamical background}.
\newblock {\em Phys. Rev. D}, 81(8):084047, 2010.

\bibitem[BVdBM62]{SeriesVII}
H.~Bondi, M.~G.~J Van~der Burg, and A.~W.~K. Metzner.
\newblock {Gravitational waves in general relativity, VII. Waves from
  axi-symmetric isolated system}.
\newblock {\em Proc. R. Soc. A}, 269(1336):21--52, 1962.

\bibitem[CGZ16]{zimmerman1}
M.~Casals, S.~E. Gralla, and P.~Zimmerman.
\newblock Horizon instability of extremal {K}err black holes: Nonaxisymmetric
  modes and enhanced growth rate.
\newblock {\em Phys. Rev. D}, 94:064003, 2016.

\bibitem[Chr02]{Chr02}
D.~Christodoulou.
\newblock {The Global Initial Value Problem in General Relativity}.
\newblock In {\em The Ninth Marcel Grossmann Meeting}, pages 44--54. World
  Scientific Publishing Company, 2002.

\bibitem[CK93]{CK93}
D.~Christodoulou and S.~Klainerman.
\newblock {\em {The Global Nonlinear Stability of the Minkowski Space}}.
\newblock vol. 41 of Princeton Mathematical Series, Princeton University Press,
  1993.

\bibitem[CT84]{CouchTorrence84}
W.~Couch and R.~Torrence.
\newblock {Conformal invariance under spatial inversion of extreme
  Reissner-Nordstr\"om black holes}.
\newblock {\em General Relativity and Gravitation}, 16(8):789--792, August
  1984.

\bibitem[Daf05]{Dafermos05interior}
M.~Dafermos.
\newblock The interior of charged black holes and the problem of uniqueness in
  general relativity.
\newblock {\em Commun. Pure Appl. Math.}, LVIII:0445--0504, 2005.

\bibitem[Dam86]{Damour86}
T.~Damour.
\newblock {Analytical calculations of gravitational radiation}.
\newblock In {\em The Fourth Marcel Grossmann Meeting}, pages 365--392.
  Elsevier Science Publishers, 1986.

\bibitem[DHRT21]{DHRT21}
Mihalis Dafermos, Gustav Holzegel, Igor Rodnianski, and Martin Taylor.
\newblock {The non-linear stability of the Schwarzschild family of black
  holes}.
\newblock {\em arXiv:2104.08222}, 2021.

\bibitem[DL17]{DafermosLuk17}
M.~Dafermos and J.~Luk.
\newblock The interior of dynamical vacuum black holes {I}: The
  {$C^0$-stability} of the {K}err {C}auchy horizon.
\newblock {\em arXiv:1710.01722}, 2017.

\bibitem[DRS18]{DRSR18}
Mihalis {Dafermos}, Igor {Rodnianski}, and Yakov {Shlapentokh-Rothman}.
\newblock {A scattering theory for the wave equation on Kerr black hole
  exteriors}.
\newblock {\em {Ann. Sci. Éc. Norm. Supér.}}, 51:371--486, April 2018.

\bibitem[DRSR16]{DRSR16}
M.~Dafermos, I.~Rodnianski, and Y.~Shlapentokh-Rothman.
\newblock Decay for solutions of the wave equation on {K}err exterior
  spacetimes {III: The full subextremal case} $|a| < m$.
\newblock {\em Annals of Math.}, 183:787--913, 2016.

\bibitem[GA01]{glampedakisfull}
K.~Glampedakis and N.~Andersson.
\newblock Late-time dynamics of rapidly rotating black holes.
\newblock {\em Phys. Rev. D}, 64:104021, 2001.

\bibitem[Gaj21]{Gajic21Oberwolfach}
D.~Gajic.
\newblock {Azimuthal instabilities of extremal black holes}.
\newblock {\em Oberwolfach Workshop Reports}, 40, 2021.

\bibitem[Gaj22]{Gajic22Inverse}
D.~Gajic.
\newblock {Late-time asymptotics for wave equations with inverse-square
  potentials}.
\newblock {\em arXiv:2203.15838}, 2022.

\bibitem[Gaj23]{Gajic23Kerr}
D.~Gajic.
\newblock {Azimuthal instabilities on extremal Kerr}.
\newblock {\em arXiv:2302.06636}, 2023.

\bibitem[Hin22]{Hintz22}
Peter Hintz.
\newblock {A Sharp Version of Price’s Law for Wave Decay on Asymptotically
  Flat Spacetimes}.
\newblock {\em Comm. Math. Physics}, 389:491--542, 2022.

\bibitem[{Keh}21a]{I}
L.~M.~A. {Kehrberger}.
\newblock {The Case Against Smooth Null Infinity I: Heuristics and
  Counter-Examples}.
\newblock {\em Ann. Henri Poincaré}, 23:829--921, 2021.

\bibitem[{Keh}21b]{II}
L.~M.~A. {Kehrberger}.
\newblock {The Case Against Smooth Null Infinity II: A Logarithmically Modified
  Price's Law}.
\newblock {\em arXiv:2105.08084}, 2021.
\newblock [Accepted in Adv. Theor. Math. Phys.].

\bibitem[{Keh}22]{III}
L.~M.~A. {Kehrberger}.
\newblock {The Case Against Smooth Null Infinity III: Early-Time Asymptotics
  for Higher $\ell$-Modes of Linear Waves on a Schwarzschild Background}.
\newblock {\em Ann. PDE}, 8(12), 2022.

\bibitem[{Keh}23]{IV}
L.~M.~A. {Kehrberger}.
\newblock {The Case Against Smooth Null Infinity IV: Early-Time Asymptotics for
  Linearised Gravity Around Schwarzschild--An Overview}.
\newblock {\em to appear in Phil. Trans. Roy. Soc. A}, 2023.

\bibitem[KM23]{V}
L.~M.~A. {Kehrberger} and H.~Masaood.
\newblock {The Case Against Smooth Null Infinity V: Early-Time Asymptotics for
  Linearised Gravity Around Schwarzschild--A Fixed-Frequency Analysis (working
  title)}.
\newblock {\em to appear}, 2023.

\bibitem[Kro00]{Kroon00}
J.~A.~V. Kroon.
\newblock {Polyhomogeneity and zero-rest-mass fields with applications to
  Newman-Penrose constants}.
\newblock {\em Class. Quantum Grav.}, 17(3):605--621, 2000.

\bibitem[Kro01]{Kroon01}
J.~A.~V. Kroon.
\newblock {Can one detect a non-smooth null infinity?}
\newblock {\em Class. Quantum Grav.}, 18(20):4311--4316, 2001.

\bibitem[KS21]{KlSz21}
S.~Klainerman and J.~Szeftel.
\newblock {Kerr stability for small angular momentum}.
\newblock {\em arXiv:2104.11857}, 2021.

\bibitem[Lea86]{Leaver86}
E.~W. Leaver.
\newblock {Spectral decomposition of the perturbation response of the
  Schwarzschild geometry}.
\newblock {\em Phys. Rev. D}, 34:384--408, 1986.

\bibitem[LMRT13]{Lucietti2013}
James {Lucietti}, Keiju {Murata}, Harvey~S. {Reall}, and Norihiro {Tanahashi}.
\newblock {On the horizon instability of an extreme Reissner-Nordstr{\"o}m
  black hole}.
\newblock {\em Journal of High Energy Physics}, 2013:35(3):1--44, 2013.

\bibitem[LO19]{LukOh16}
J.~Luk and S.-J. Oh.
\newblock {Strong cosmic censorship in spherical symmetry for two-ended
  asymptotically flat data I: Interior of the black hole region}.
\newblock {\em Annals of Math.}, 190(1):1--111, 2019.

\bibitem[Luk21]{Luktalk21}
J.~Luk.
\newblock {A tale of two tails}, October 2021.
\newblock {Talk at IPAM Workshop II: Mathematical and Numerical Aspects of
  Gravitation, \url{https://mathinstitutes.org/videos/18141}}.

\bibitem[Mas22]{Masaood22}
H.~Masaood.
\newblock {A Scattering Theory for Linearised Gravity on the Exterior of the
  Schwarzschild Black Hole I: The Teukolsky Equations}.
\newblock {\em Commun. Math. Phys.}, 393:477–581, 2022.

\bibitem[MZ22]{MZ21}
Siyuan Ma and Lin Zhang.
\newblock Price’s law for spin fields on a schwarzschild background.
\newblock {\em Annals of PDE}, 8, 11 2022.

\bibitem[MZ23]{MZ23}
S.~Ma and L.~Zhang.
\newblock {Sharp Decay for Teukolsky Equation in Kerr Spacetimes}.
\newblock {\em Comm. Math. Phys.}, 2023.

\bibitem[Nic16]{nicolas}
J.~P. Nicolas.
\newblock {Conformal scattering on the Schwarzschild metric}.
\newblock {\em Annales de l'Institut Fourier}, 66(3):1175--1216, 2016.

\bibitem[NP62]{NP62Approach}
E.~T. Newman and R.~Penrose.
\newblock An approach to gravitational radiation by a method of spin
  coefficients.
\newblock {\em J. Math. Phys.}, 3:566--768, 1962.

\bibitem[NP65]{NPconstants65}
E.~T. Newman and R.~Penrose.
\newblock {10 Exact Gravitationally-Conserved Quantities}.
\newblock {\em Phys. Rev. Lett.}, 15:231--233, 1965.

\bibitem[NP68]{NPconstants68}
E.~T. Newman and R.~Penrose.
\newblock {New conservation laws for zero rest-mass fields in asymptotically
  flat space-time}.
\newblock {\em Proc. R. Soc. A}, 305(1481):175--204, 1968.

\bibitem[Ori13]{ori2013}
A.~Ori.
\newblock Late-time tails in extremal {R}eissner--{N}ordstr\"{o}m spacetime.
\newblock {\em arXiv:1305.1564}, 2013.

\bibitem[Pen65]{Penrose65}
R.~Penrose.
\newblock {Zero rest-mass fields including gravitation: asymptotic behaviour}.
\newblock {\em Proc. R. Soc. A}, 284(1397):159--203, 1965.

\bibitem[Pen69]{Penrose69collapse}
R.~Penrose.
\newblock Gravitational collapse: the role of general relativity.
\newblock {\em Rev. del Nuovo Cimento}, 1:272--276, 1969.

\bibitem[Pri72]{Price72}
R.~Price.
\newblock Nonspherical perturbations of relativistic gravitational collapse.
  {I}. scalar and gravitational perturbations.
\newblock {\em Phys. Rev. D}, 5:2419--2438, 1972.

\bibitem[Sac61]{SeriesVI}
R.~Sachs.
\newblock {Gravitational waves in general relativity VI. The outgoing radiation
  condition}.
\newblock {\em Proc. R. Soc. A}, 264(1318):309--338, 1961.

\bibitem[Sac62]{SeriesVIII}
R.~Sachs.
\newblock {Gravitational waves in general relativity VIII. Waves in
  asymptotically flat space-time}.
\newblock {\em Proc. R. Soc. A}, 270(1340):103--126, 1962.

\bibitem[Sel16]{sela}
O.~Sela.
\newblock Late-time decay of perturbations outside extremal charged black hole.
\newblock {\em Phys. Rev. D}, 93:024054, 2016.

\bibitem[SRTdC20]{SRTdC20}
Y.~Shlapentokh-Rothman and R.~Teixeira~da Costa.
\newblock {Boundedness and decay for the Teukolsky equation on Kerr in the full
  subextremal range $| a|< M $: frequency space analysis}.
\newblock {\em arXiv:2007.07211}, 2020.

\bibitem[SRTdC23]{SRTdC23}
Y.~Shlapentokh-Rothman and R.~Teixeira~da Costa.
\newblock {Boundedness and decay for the Teukolsky equation on Kerr in the full
  subextremal range |a|<M: physical space analysis}.
\newblock {\em arXiv:2302.08916}, 2023.

\bibitem[{Teu}73]{Teukolsky73}
S.~A. {Teukolsky}.
\newblock {Perturbations of a Rotating Black Hole. I. Fundamental Equations for
  Gravitational, Electromagnetic, and Neutrino-Field Perturbations}.
\newblock {\em APJ}, 185:635--648, 1973.

\bibitem[TP74]{Teukolsky74}
S.~A. {Teukolsky} and W.~H. {Press}.
\newblock {Perturbations of a Rotating Black Bole. III. Interaction of the Hole
  With Gravitational and Electromagnetic Radiation}.
\newblock {\em APJ}, 193:443--461, 1974.

\bibitem[Wal73]{WaldTeuk}
R.~M. Wald.
\newblock {On perturbations of a Kerr black hole}.
\newblock {\em J. Math. Phys.}, 14(10):1453--1461, 1973.

\bibitem[WW79]{WalkerWill79}
M.~Walker and C.~M. Will.
\newblock {Relativistic Kepler problem. II. Asymptotic behavior of the field in
  the infinite past}.
\newblock {\em Phys. Rev. D}, 19(12):3495--3508, 1979.

\bibitem[YZZ{\etalchar{+}}13]{zengi13}
H.~Yang, A.~Zimmerman, A.~Zenginoglu, F.~Zhang, E.~Berti, and Y.~Chen.
\newblock Quasinormal modes of nearly extremal {K}err spacetimes: {S}pectrum
  bifurcation and power-law ringdown.
\newblock {\em Phys. Rev. D}, 88:044047, 2013.

\bibitem[ZKB14]{zengi14}
A.~Zengino{\u{g}}lu, G.~Khanna, and L.~M. Burko.
\newblock {Intermediate behavior of {K}err tails}.
\newblock {\em Gen. Rel. Grav.}, 46(3):1672, 2014.

\end{thebibliography}
\end{document}